\documentclass{aa} 
\usepackage{graphicx}
\usepackage{lipsum}
\usepackage[varg]{txfonts}
\usepackage{natbib}
\bibpunct{(}{)}{;}{a}{}{,} 
\usepackage{url}
\usepackage{hyperref}
\usepackage{amsmath,bm}
\usepackage{color}

\definecolor{darkcyan}{RGB}{0,50,250}

\begin{document}

\title{The coronal mass ejection productivity of solar active region 13664/8} 
\titlerunning{The coronal mass ejection productivity of solar active region 13664/8} 

\author{Lijuan Liu\inst{1,2,3}\fnmsep\thanks{Corresponding author: liulj8@mail.sysu.edu.cn}
     \and Yuming Wang\inst{4,2}\fnmsep\thanks{Corresponding author: ymwang@ustc.edu.cn}
     \and Quanhao Zhang\inst{4,2}
     \and Jingnan Guo\inst{4,2}
     \and Yutian Chi\inst{5}}
\authorrunning{Lijuan Liu et al.}
\institute{Planetary Environmental and Astrobiological Research Laboratory (PEARL), School of Atmospheric Sciences, Sun Yat-sen University, Zhuhai, Guangdong, 519082, China
  \and CAS center for Excellence in Comparative Planetology, China
  \and Key Laboratory of Tropical Atmosphere-Ocean System, Sun Yat-sen University, Ministry of Education, Zhuhai, China
  \and CAS Key Laboratory of Geospace Environment, Department of Geophysics and Planetary Sciences, University of Science and Technology of China, Hefei, Anhui, 230026, China
  \and Institute of Deep Space Sciences, Deep Space Exploration Laboratory, Hefei 230088, China}
  
\date{Received 22 December 2025 / Accepted 14 July 2026} 

  
\abstract{In May 2024, NOAA active region (AR) complex 13664/8 
appeared as one of the most productive regions of the current solar cycle. 
It produced 12 X-class flares and over 20 coronal mass ejections (CMEs), 
triggering the strongest geomagnetic storm since 2003.} 
{We investigated why the AR complex is so productive, particularly in CMEs.} 
{Primarily using observations from the Atmospheric Imaging Assembly and Helioseismic and Magnetic Imager aboard the Solar Dynamics Observatory, 
we analyzed the photospheric magnetic evolution, eruption sources, and eruption waiting times and compared the magnetic parameters with five other ARs.} 
{The region initially comprised only AR 13664 and exhibited limited flare activity until AR 13668 emerged on May 4, 
after which clustered major flares and CMEs occurred. 
Rapid, complex flux emergence substantially increased the region's area, magnetic flux, complexity, and non-potentiality.    
A comparison with five other ARs suggests that both flare-rich and CME-rich ARs exhibit elevated overall non-potentiality, 
as indicated by their high total magnetic free energy density and total current helicity, while CME-rich ARs exhibit particularly large mean current helicity, possibly implying stronger localized non-potentiality.   
The increased complexity manifests through at least 12 emerging bipoles and six collisional polarity inversion lines (cPILs) formed between nonconjugated polarities.    
All six cPILs exhibit sustained collision and shearing, serving as eruption sources. 
Decay index distributions show systematically lower critical heights ($<45$ Mm) for a torus instability above the CME sources. 
The CME waiting time distribution exhibits two peaks, 
suggesting that multiple cPILs enhance CME productivity by increasing both recurrent CMEs from the same location and disturbance-triggered CMEs from nearby locations.}
{The results highlight that in addition to sufficient non-potentiality and rapid background field decay, 
a high degree of magnetic complexity accompanied by 
dynamical collisional shearing at multiple cPILs is crucial for the region's extreme CME productivity.}


\maketitle
\nolinenumbers

\section{Introduction}\label{sec:intro}

Solar active regions (ARs), 
localized concentrations of strong magnetic fields on the Sun, are the primary sources of solar flares and coronal mass ejections (CMEs)~\citep{Van_2015},  
both of which are major drivers of hazardous space weather~\citep[e.g.,][]{Webb_2012}. 
While most ARs produce only minor flares or sporadic major events~\citep{Wangym_2011,Chen_Wang_2011}, a subset---often referred to as super ARs~\citep[SARs; e.g.,][]{Bai_1988, Tian_2002, Chen_Wang_2012}---exhibits exceptional eruptivity by unleashing clusters of major flares and CMEs over days. 
These productive ARs are usually magnetically complex,  
with configurations far more intricate than simple bipolar regions~\citep[e.g.,][]{Toriumi_2019}. 
The widely used Hale (Mount Wilson) classification~\citep{Hale_1919, Kunzel_1960} describes AR complexity from the morphological arrangement of magnetic polarities,   
distinguishing unipolar and bipolar regions ($\alpha$ and $\beta$), 
mixed or irregular polarity distributions ($\gamma$ and $\beta\gamma$), and $\delta$ sunspots in which umbrae of opposite polarities share a common penumbra.   
Statistical surveys suggest that more than $80\%$ of productive ARs contain $\delta$ sunspots---the highest complexity level in the Hale classification system~\citep{Toriumi_2016}.

From a magnetic perspective, such morphologically complex ARs often exhibit strong magnetic non-potentiality, 
manifested by extended shear lengths, stronger magnetic gradients, 
larger electric currents, magnetic free energy, and magnetic helicity. 
These quantities have been shown to correlate with flare and CME productivity in many statistical and case studies~\citep[e.g.,][]{Falconer_2002, Leka_Barnes_2003a, Jing_2006, Falconer_etal_2006,  Leka_Barnes_2007, Schrijver_2007, Wang_Zhang_2008, Tziotziou_2012, Tziotziou_2013, Bobra_Couvidat_2015, Bobra_Ilonidis_2016, Toriumi_2016, Sun_2015, Liu_2016, Vemareddy_2019}.

While both flares and CMEs are linked to non-potentiality, 
successful CME eruptions additionally require relatively weak confinement by the overlying magnetic field~\citep{Kliem_2006, Wangym_2007}. 
This confinement is commonly assessed using the decay index, which measures how rapidly the external field decreases with height~\citep{Kliem_2006}. 
In the ideal torus instability theory, 
a successful eruption occurs once the magnetic flux rope---a central structure in the standard flare and CME model~\citep[see review in][]{Shibata_2011}---reaches heights where the decay index exceeds the critical threshold of approximately 1.5~\citep{Kliem_2006}. 
Case and statistical studies further confirm that confined events are typically associated with higher critical heights~\citep[e.g.,][]{Liuy_2008, Guo_2010, Chengx_2011a, Sun_2015, Liu_2016, Wangd_2017, Baumgartner_2018, Filippov_2020, James_2022, Gupta_2024, Yang_2024}, 
with empirical dividing heights between eruptive and confined events often found around 40 Mm--50 Mm~\citep[e.g.,][]{Wangd_2017, Baumgartner_2018}.

At smaller scales, 
the magnetic sources of major flares and CMEs are commonly traced to polarity inversion lines (PILs), 
particularly those that are strongly sheared and exhibit high magnetic gradients between strong opposite polarities~\citep[e.g.,][]{Schrijver_2007, Schrijver_2009, Toriumi_2019, Dhakal_2024}. 
Such PILs are frequently linked to $\delta$ sunspots~\citep{Toriumi_2016, Toriumi_2017, Chintzoglou_2018}.  
The recently proposed collisional shearing scenario, 
which emphasizes the AR evolution rather than static properties, 
explains the formation of such PILs. 
In this scenario, the natural separation of conjugated polarities during the flux emergence can bring nonconjugated, opposite-signed polarities into collision. Persistent collision leads to shearing and flux cancellation along the PIL, 
creating conditions favorable for magnetic reconnection and subsequent flares and CMEs.  
Polarity inversion lines formed through this process are referred to as collisional PILs (cPILs)~\citep{Chintzoglou_2018}.

Within a single productive AR, 
multiple active PILs may develop, 
producing homologous eruptions from the same magnetic source 
and quasi-homologous eruptions from different sources~\citep{Zhang_Wang_2002, Wang_2013b, Liu_2017}. 
Successive CMEs with short waiting times can interact in the heliosphere, 
enhancing their geomagnetic effect~\citep[e.g.,][]{Liuyd_2014, Chi_2016, Shen_2018, Camilla_2020, Shen_2021}. 
Homologous or quasi-homologous CMEs are therefore potential contributors to intense geomagnetic storms, 
further highlighting the role of highly productive ARs in driving extreme space weather events. 
This raises a fundamental question: 
what intrinsic properties enable certain ARs to sustain high productivity,  
particularly in generating CMEs?

In May 2024, the AR complex consisting of NOAA ARs 13664 and 13668 (AR 13664/8 hereafter) appeared as one of the most productive ARs of the current solar cycle. 
Throughout its transit, 
its size became almost comparable to that of the SAR 12192, 
which is well known for producing numerous confined X-class flares in October 2014~\citep[e.g.,][]{Sun_2015, Liu_2016}. 
In contrast, AR 13664/8 produced 12 X-class flares and at least 20 major CMEs~\citep[e.g.,][]{Hayakawa_2024}, 
which triggered the most intense geomagnetic storm since the 2003 Halloween event, with the disturbance storm time (Dst) index reaching -412 nT~\citep[e.g.,][]{Sun_2024}. The AR thus offers a unique opportunity to investigate why some ARs are exceptionally CME-productive.

Numerous studies have examined the region's different aspects. 
\citet{Hayakawa_2024} provided a data report on its major events and geomagnetic effects. 
\citet{Li_2024} and \citet{Jing_2025} studied white-light flares using data from 
the Advanced Space-based Solar Observatory~\citep[ASO-S;][]{Gan_2019}, 
while \citet{Razquin_2025} investigated CME-associated coronal dimming. 
Many other studies have focused on the region's high productivity. 
\citet{Sun_2024} report an extremely fast flux emergence rate 
and emphasize the role of collisional shearing between emerging bipoles. 
Similar conclusions regarding polarity interactions are reported~\citep{Romano_2024, Kontogiannis_2024a, Jarolim_2024, Wang_2024, Jaswal_2025, MacTaggart_2025, Dikpati_2025}.  
In particular, \citet{Romano_2024} highlight the importance of shear and converging motions associated with such interactions. 
Recent studies have also reported the region's high complexity and non-potentiality~\citep{Kontogiannis_2024a, Jarolim_2024, Wang_2024, Jaswal_2025, MacTaggart_2025, Dikpati_2025, Korsos_2025}, 
while~\citet{Jarolim_2024} and~\citet{MacTaggart_2025} investigated its photospheric field and extrapolated coronal field evolution in detail.

Overall, the above studies suggest that the high productivity of AR 13664/8 is closely related to its rapid flux emergence, strong polarity interactions, and the resulting high magnetic complexity and non-potentiality. 
However, few studies have explicitly addressed why this AR is exceptionally productive in CMEs. 
To address this question, 
we investigated the magnetic evolution of the AR complex, 
the precise magnetic sources of its major flares and CMEs, and waiting times between events.

\section{Data and methodology}\label{sec:data} 

\begin{figure*}
\begin{center}
\includegraphics[width=1.02\hsize]{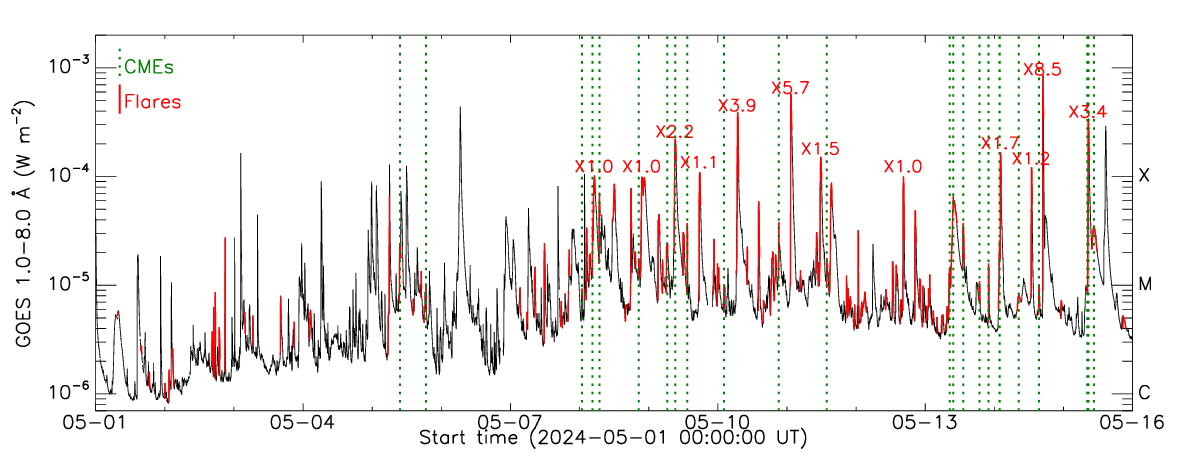} 
\caption{
GOES soft X-ray light curve during the disk transit of AR 13664/8. The superimposed red lines denote all flares originating from the AR complex, while the vertical green lines specifically indicate the start times of the CME-associated flares. 
}\label{fig:goes} 
\end{center}
\end{figure*}

During its disk transit, 
AR 13664/8 produced 120 flares above C-class, including 59 M-class and 12 X-class flares. 
Among them, 23 were associated with CMEs (Fig.~\ref{fig:goes}). 
We studied the 25 most significant events, including 23 CMEs associated with C- to X-class flares and two confined X-class flares (Table~\ref{tb:eru}). 
Among the 25 cases, 
12 CMEs and two confined X-class flares (events 1-14 in Table~\ref{tb:eru}) occurred within Stonyhurst longitudes 75$^\circ$E--75$^\circ$W, 
enabling the precise identification of their magnetic sources with relatively 
high-resolution, high-cadence data from Earth-viewing satellites. 
The data and methods are described below.

\begin{table}
\caption{25 most significant events from AR 13664/8.}\label{tb:eru}
\centering
\resizebox{\columnwidth}{!}{%
\footnotesize
\begin{tabular}{l|cccc|c|c}
\hline
No. & \multicolumn{4}{c|}{\centering{Flare}} & CME association & Magnetic source \\
\hline
 & Start & End & Peak & Class & &  \\
\hline
1 & 2024-05-05 14:12:00 & 14:26:00 & 14:16:00 & C7.5  & Y & PIL1 \\
2 & 2024-05-07 03:18:00 & 03:33:00 & 03:20:00 & C9.7  & Y & PIL2 \\
3 & 2024-05-08 02:16:00 & 02:36:00 & 02:27:00 & M3.4  & Y & PIL3  \\
4 & 2024-05-08 04:37:00 & 05:28:20 & 05:09:00 & X1.0  & Y & PIL2 \\
5 & 2024-05-08 11:26:00 & 12:22:18 & 12:04:00 & M8.7  & Y & PIL4 \\
6 & 2024-05-08 21:08:00 & 21:58:17 & 21:40:00 & X1.0  & Y & PIL4 \\
7 & 2024-05-09 08:45:00 & 09:31:20 & 09:13:00 & X2.3  & Y & PIL4 \\
8 & 2024-05-09 11:52:00 & 12:02:00 & 11:56:00 & M3.1  & Y & PIL5 \\
9 & 2024-05-09 17:23:00 & 18:01:19 & 17:44:00 & X1.1  & Y & PIL3 \\
10 &  2024-05-10 06:27:00 & 07:13:19 & 06:54:00 & X4.0  & Y & PIL3 \\
11 &  2024-05-11 01:10:00 & 01:39:00 & 01:23:00 & X5.8  & Y & PIL3 \\
12 &  2024-05-11 11:15:00 & 12:01:20 & 11:44:00 & X1.5  & N & PIL1 \\
13 &  2024-05-11 14:46:00 & 15:43:20 & 15:25:00 & M8.9  & Y & PIL3 \\
14 &  2024-05-12 16:11:00 & 16:43:24 & 16:26:00 & X1.0  & N & PIL6 \\
15 &  2024-05-13 08:48:00 & 10:57:00 & 09:44:00 & M6.6  & Y & \\
16\hyperlink{target1}{$^*$} &  2024-05-13 12:56:00 & 13:28:21 & 13:11:00 & M3.7  & Y & \\
17\hyperlink{target1}{$^*$} &  2024-05-13 18:37:00 & 19:02:00 & 18:48:00 & M1.1  & Y & \\
18\hyperlink{target1}{$^*$} &  2024-05-13 21:48:00 & 22:07:00 & 21:59:00 & M1.5  & Y & \\
19 &  2024-05-14 01:23:00 & 02:07:18 & 01:48:00 & M2.6  & Y & \\
20 &  2024-05-14 02:03:00 & 02:19:00 & 02:09:00 & X1.7  & Y & \\
21 &  2024-05-14 12:40:00 & 13:05:00 & 12:55:00 & X1.2  & Y & \\
22 &  2024-05-14 16:46:00 & 17:02:00 & 16:51:00 & X8.7  & Y & \\
23 &  2024-05-15 08:18:00 & 08:55:34 & 08:37:00 & X3.5  & Y & \\
24 &  2024-05-15 09:47:00 & 11:40:00 & 10:35:00 & M3.6  & Y & \\
25 &  2024-05-15 20:30:00 & 21:43:00 & 20:39:00 & C5.2  & Y & \\

\hline  
\end{tabular} 
}
\tablefoot{All times are given in UT. \\ 
\hypertarget{target1}{$^*$} For cases 16--18, no associated CMEs were recorded in the SOHO LASCO CME Catalog.  
However, narrow, jet-like eruptions were discernible in LASCO/C2 images. We nonetheless classify these flares as CME-associated here.} 
\end{table}

\subsection{Data}\label{subsec:DM:data}
We used GOES-18 soft X-ray flux, multiwavelength images from the Atmospheric Imaging Assembly~\citep[AIA,][]{Lemen_etal_2012} aboard the Solar Dynamics Observatory~\citep[SDO,][]{Pesnell_2012}, 
photospheric magnetic field from the Helioseismic and Magnetic Imager~\citep[HMI,][]{Scherrer_2012} aboard SDO, and coronagraphic images from the Large Angle and Spectrometric Coronagraph (LASCO) C2 aboard the Solar and Heliospheric Observatory ~\citep[SOHO,][]{Domingo_2000}. 
The ultraviolet (UV) and extreme-ultraviolet (EUV) images from SDO/AIA have a plate scale of 0\farcs 6 and a cadence up to 12~s. 
For the photospheric magnetic field, we used the Space-weather HMI Active Region Patches~\citep[SHARPs;][]{Hoeksema_etal_2014, Bobra_2014}, 
which provide automatically tracked AR cutouts of line-of-sight and vector magnetic field with a plate scale of 0\farcs 5 and a cadence of 720 s. 
Specifically, we used the SHARP data series \texttt{hmi.sharp\_cea\_720s}, 
remapped from charge-coupled device (CCD) coordinates to cylindrical equal-area (CEA) heliographic coordinates.

\subsection{Analysis of photospheric magnetic field evolution}\label{subsec:DM:evo-mag} 
To track the main polarity evolution, 
we first identified the local field-strength maximum of each polarity in each $B_z$ (vertical component) magnetogram 
and then computed the flux-weighted centroid within a circular region of radius $\sim$5 Mm centered on that maximum~\citep{Chintzoglou_2018, Liu_2021}.  
The resulting polarity evolution is shown in Sect.~\ref{subsubsec:field} (Fig.~\ref{fig:bz}).

We further examined the temporal evolution of several photospheric SHARP parameters, 
including unsigned magnetic flux $\Phi$, a proxy for total photospheric magnetic free energy density $\rho_{total}$, total current helicity ${H_c}_{total}$ ($B_z$ contribution), and mean current helicity $\overline{H_c}$ ($B_z$ contribution).   
The parameters were computed by $\Phi=\Sigma|B_z|dA $, $\displaystyle \rho_{total} = \Sigma \frac{1}{8 \pi} (B_{obs}-B_{pot})^2dA$, ${H_c}_{total} = {\Sigma | B_z (\nabla \times B)_z | } $, and $\displaystyle  \overline{H_c} = \frac{1 }{N} \Sigma B_z (\nabla \times B)_z $, 
where $B_{obs}$ is the observed vector magnetic field, $B_{pot}$ is the extrapolated potential field (see extrapolation procedure below), $dA$ is the pixel area, and $N$ is the number of pixels.

While $\Phi$ provides an overall measure of the AR's magnetic flux, 
the other three parameters quantify its non-potentiality. 
The $\rho_{total}$ serves as a photospheric proxy for the AR's free magnetic energy, 
while the current helicity parameters ${H_c}_{total}$ and $\overline{H_c}$ were used as proxies for magnetic helicity in a nearly force-free magnetic field where the field and current are approximately parallel. 
These parameters together characterize complementary physical aspects of ARs 
and have been shown to help distinguish flare and CME productivity among different ARs~\citep{Liu_2016}. 
We therefore adopted these four parameters to characterize the AR complex and directly compare AR 13664/8 with five other ARs analyzed in~\citet{Liu_2016}, 
including the flare-rich but CME-poor AR 12192. 
The parameter evolution is shown in Sect.~\ref{subsubsec:para} (Fig.~\ref{fig:para}).

\subsection{Flare identification and flare–CME association}\label{subsec:DM:flare-CME}
To identify all flares from the AR complex, 
we first combined the flare lists released by the NOAA Space Weather Prediction Center\footnote{\url{ftp://ftp.swpc.noaa.gov/pub/warehouse/}} and SolarMonitor\footnote{\url{https://www.solarmonitor.org/}} 
and then manually inspected the corresponding AIA 131~\AA~images. 
A flare was attributed to AR 13664/8 if continuous flaring occurred within the AR, 
its timing matched the GOES soft X-ray light curve evolution, 
and no brighter flaring occurred elsewhere. 
An identification example is provided in Sect.~\ref{subsubsec:source} (Fig.~\ref{fig:source}a-b).

We examined the CME association for each flare 
by combining eruption evidence in both the low and high corona. 
First, we inspected the AIA EUV images for on-disk and low-coronal eruption signatures, 
such as filament, hot channel,
or cavity eruptions, EUV waves, and coronal dimming~\citep[e.g.,][]{wangym_2002, Wangym_2011}. 
Second, we checked the LASCO/C2 observations, 
using the SOHO LASCO CME catalog \footnote{\label{fn:cme_catalog}\url{https://cdaw.gsfc.nasa.gov/CME_list/index.html}} 
as a reference. 
A CME was considered associated with a flare that showed observable on-disk or low-coronal signatures if it first appeared in LASCO/C2 within approximately $\pm 1$ hour of the flare peak time and its propagation direction was consistent with the AR disk location. 
We also examined eruptions occurring near the flare time but originating elsewhere 
to avoid misidentification. 
Finally, we cross-checked our CME list with that reported by~\citet{Hayakawa_2024}. 
Except for the three narrow CMEs that occurred on May 13 (case 16--18 in Table~\ref{tb:eru}), 
our CME list is consistent with theirs. 
An association example is provided in Sect.~\ref{subsubsec:source} (Fig.~\ref{fig:source}b-c).

\subsection{Identification of magnetic sources for major flares and CMEs}\label{subsec:DM:mag-source}
We combined the SHARP $B_z$ maps with EUV and UV images to identify the precise magnetic sources of the 14 major events (events 1-14 in Table~\ref{tb:eru}). 
We first remapped the AIA images to the same CEA heliographic coordinates as the SHARP maps 
and then created composite images by overlaying the two datasets. 
In each composite, 
the 1600 \AA\ channel and one EUV channel 
(e.g., 131 \AA) were used. 
The former traces flare ribbons, while the latter reveals flare loops and other coronal flare structures. 
By following localized flare features---such as flare kernels, flare ribbons, and post-flare loops---overlaid on the $B_z$ maps, 
we pinpointed the events' precise magnetic sources. 
An identification example is shown in Sect.~\ref{subsubsec:source} (see Fig.~\ref{fig:source}d-f). The sources of all 14 events are shown in Fig.~\ref{fig:loc}.

Projection effects may affect source identification for limb events. 
In our sample, only event 14 occurred near 75$^\circ$W. 
Its source was determined using additional constraints from the relative positions of flare ribbons and sunspots. 
Flare ribbons form low in the solar atmosphere 
and are less affected by line-of-sight projection, 
while sunspots are readily associated with the main polarities. 
The other events occurred within 60$^\circ$E--60$^\circ$W  
and were therefore not significantly affected by projection effects.

For each event, 
we extracted its source PIL from the pre-event magnetogram by tracing the $B_z=0$ contours, 
supplemented by flare signatures such as ribbon elongation. 
These PILs were used to calculate the decay index. 
The polarity evolution forming the PILs is presented in Sect.~\ref{subsubsec:pil} (Figs.~\ref{fig:PIL1}-\ref{fig:PIL6}).

\subsection{Decay index calculation}\label{subsec:DM:decay}
The coronal potential magnetic field was extrapolated to calculate the decay index and derive the photospheric magnetic parameters.  
We performed the extrapolation using the fast Fourier transform (FFT), 
with the photospheric $B_z$ as the boundary condition. 
In a potential field, the current-free condition gives $B=\nabla \phi$, 
where $\phi$ is the scalar magnetic potential. 
Combined with the divergence-free condition $\nabla \cdot B=0$, 
this leads to the Laplace equation $\nabla^2 \phi=0$. 
We applied the FFT to $B_z$, 
solved the Laplace equation analytically in Fourier space, 
and then performed an inverse transform to obtain $\phi$ and the corresponding potential $B$ in physical space~\citep[e.g.,][]{Alissandrakis_1981, Gary_1989}. 
The extrapolation was carried out in a box of size $(601, 256, 200)$ with a grid spacing of 0.73~Mm (twice the plate scale of SHARP data).

Based on the extrapolated potential field, 
we calculated the decay index above the 14 identified source PILs, 
using $\displaystyle n=-\frac{d\ln B_{ex}(h)}{d\ln h}$. 
The decay index was averaged along each PIL as a function of height. 
The results are shown in Sect.~\ref{subsubsec:decay} (Fig.~\ref{fig:decay}).

\subsection{Waiting time analysis}\label{subsec:DM:wt} 
The waiting time of each event is defined as the time interval since its preceding event. 
Its distribution may provide information on event correlations. 
We analyzed the waiting time distributions for all flares above C-class, all X-class flares, and all CMEs. 
The results are shown in Sect.~\ref{subsec:wt} (Fig.~\ref{fig:wtime}).

\section{Results}\label{sec:res} 

\subsection{Photospheric magnetic evolution of the AR complex}\label{subsec:pho}

\subsubsection{Magnetic field evolution}\label{subsubsec:field}

\begin{figure*}
\begin{center}
\includegraphics[width=1.02\hsize]{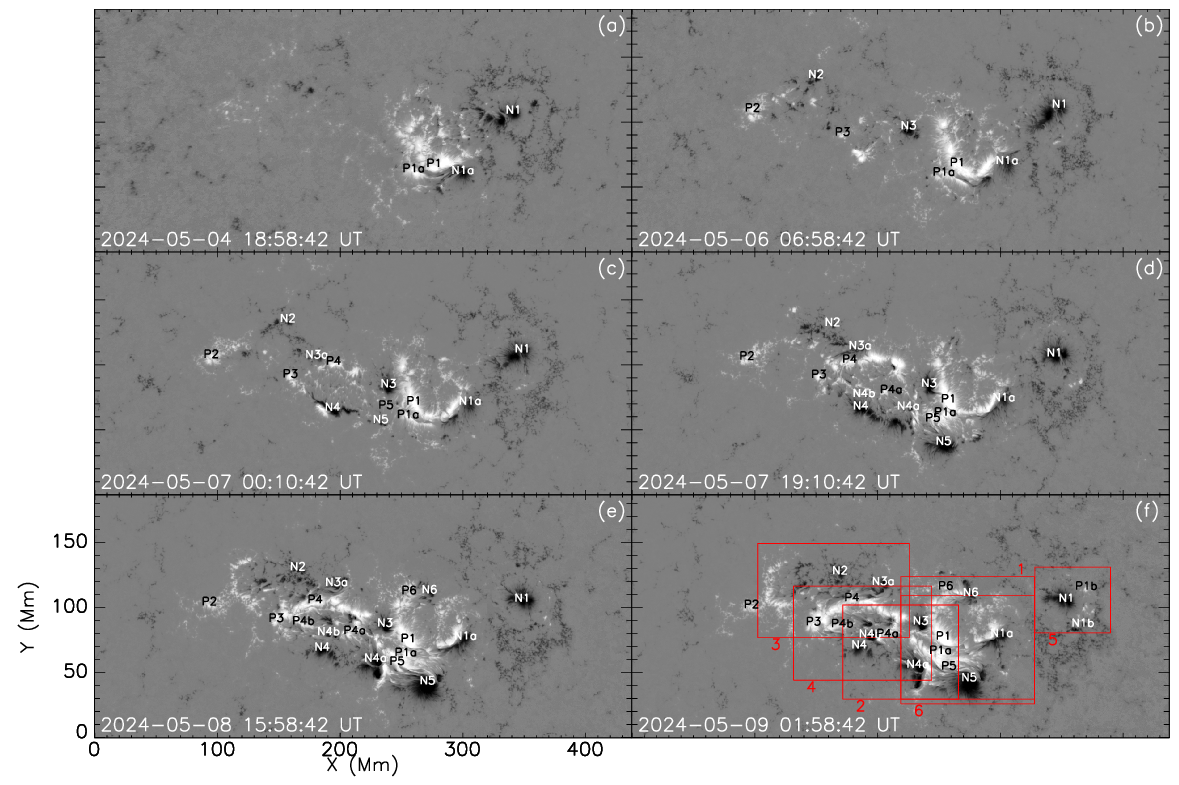}
\caption{
Evolution of the photospheric magnetic fields ($B_z$) of AR 13664/8. 
The white and black 
areas indicate positive and negative polarities, respectively, 
saturated at $\pm 2000$ Gauss. 
Flux-weighted centroids of polarities are marked by labels, 
with P1/N1 denoting the first bipole. 
The same notation is applied to subsequent polarities. 
The red boxes in panel (f) delimit the fields of view (FOVs) of Figs.~\ref{fig:PIL1}-\ref{fig:PIL6}. An animation of the $B_z$ evolution from 2024-05-02 10:58:42 UT to 2024-05-12 17:10:43 UT is available online. 
}\label{fig:bz} 
\end{center}
\end{figure*}

AR 13664 initially appeared as an overall bipolar region when it rotated into view on 2024-05-01, 
consisting of positive P1 and negative N1 polarities (Fig.~\ref{fig:bz}a), 
hereafter bipole P1-N1. 
Although phases of flux emergence were observed, 
such as bipole P1a-N1a emerging south of P1 (Fig.~\ref{fig:bz}a), 
the region largely maintained its overall bipolar configuration (see Fig.~\ref{fig:bz} associated movie) and produced only minor to moderate confined flares (no larger than M2.8 class) before May 4.

From around 22:34 UT on May 4, 
bipoles P2-N2 and P3-N3 emerged sequentially to the east of AR 13664 (Fig.~\ref{fig:bz}b), forming AR 13668. 
As their conjugated polarities separated, 
nonconjugated N3 and P1 collided, 
forming an AR complex. 
The region then entered a phase of rapid and complex flux emergence. 
On May 6, N3a, sharing P3 with N3, emerged, 
followed by two anti-Hale bipoles, P4-N4 and P5-N5, 
whose initial tilt angles were slightly opposite to the others (Fig.~\ref{fig:bz}c-d). 
Two more emergence episodes then occurred between P4 and N4, 
forming P4a-N4a and P4b-N4b (Fig.~\ref{fig:bz}d-e).   
Most newly emerged flux was concentrated within AR 13668, 
except for P6-N6 and P1b-N1b (Fig.~\ref{fig:bz}e-f). 
Except for N3a, P4a-N4a, P4b-N4b, and P1b-N1b, our polarity naming follows~\cite{Sun_2024}. 
During this phase, 
the region grew dramatically, 
with the sunspot area reaching 2761 millionths of a solar hemisphere (MSH) on May 9~\citep{Hayakawa_2024}.

Multiple episodes of complex flux emergence led to collisions among several groups of nonconjugated polarities. 
On May 9, 
when most polarity interactions were already well developed, 
several PILs exhibited clear collision signatures (Fig.~\ref{fig:bz}f). 
From east to west, 
these included PILs P3/P4/P4a/P4b-N2/N3a formed between polarity groups 
P3/P4/P4a/P4b and N2/N3a (notation used consistently hereafter),  
P3/P4/P4a/P4b-N4/N4b, P4a/P5-N4/N4a/N4b,  
P1-N3, P5-N5, P1-N1a, 
and P1b-N1 (Fig.~\ref{fig:bz}f).  
As flux emergence continued, 
additional opposite-polarity groups began to interact, 
and therefore PILs P1-N3, P5-N5, and P1-N1a evolved into P1/P6-N3/N3a, P5-N4a/N5, and P1/P5-N1a, respectively (see Fig.~\ref{fig:bz} associated movie).

In addition, the rapid emergence of multiple bipoles within a relatively restricted area led to the squeezing, shearing, and rotation of polarities along the PILs, except for P1/P6-N3/N3a (see Fig.~\ref{fig:bz} associated movie). 
Therefore, 
all the aforementioned PILs except for P1/P6-N3/N3a contained segments formed by collision and shearing between nonconjugated opposite-signed polarities  
and can be classified as cPILs~\citep{Chintzoglou_2018}.  
The detailed evolution of these PILs and their associated major flares and CMEs are shown in Sect.~\ref{subsec:source}.

\subsubsection{Magnetic parameters}\label{subsubsec:para}
\begin{figure*}[ht!]
\begin{center}
\includegraphics[width=0.49\hsize]{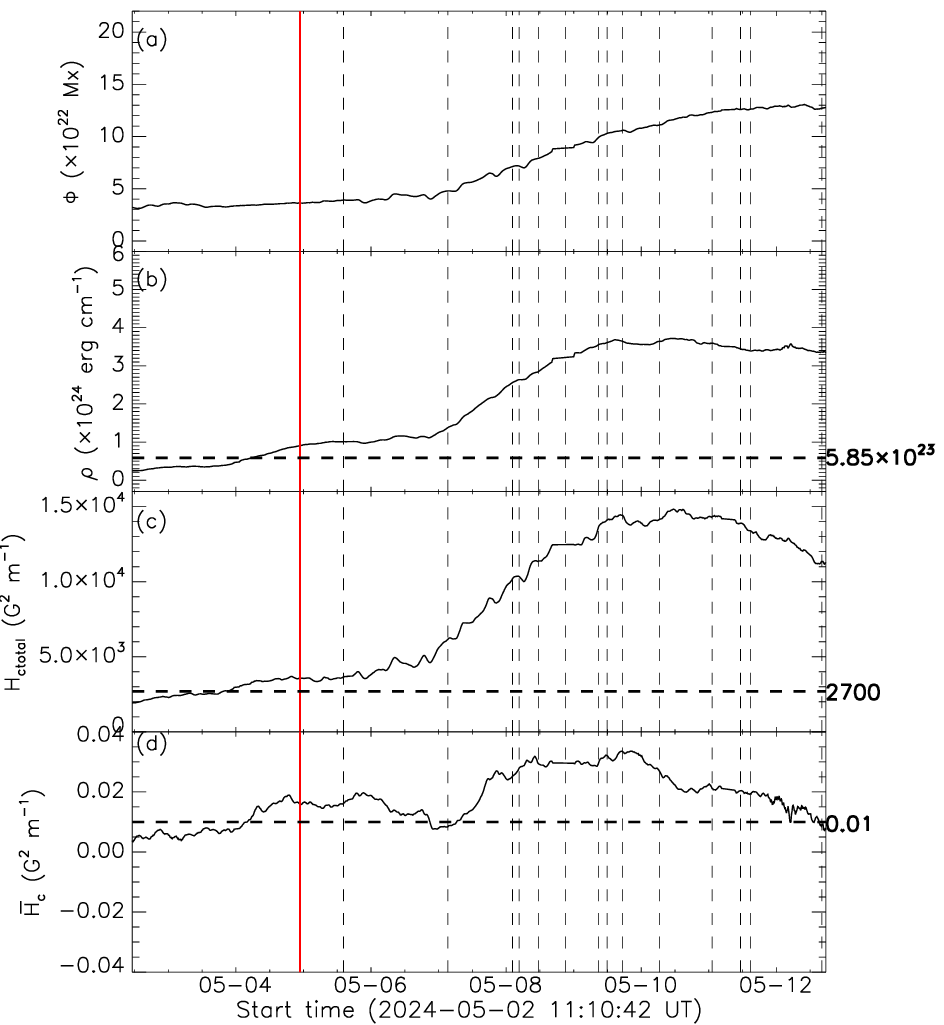}
\includegraphics[width=0.49\hsize]{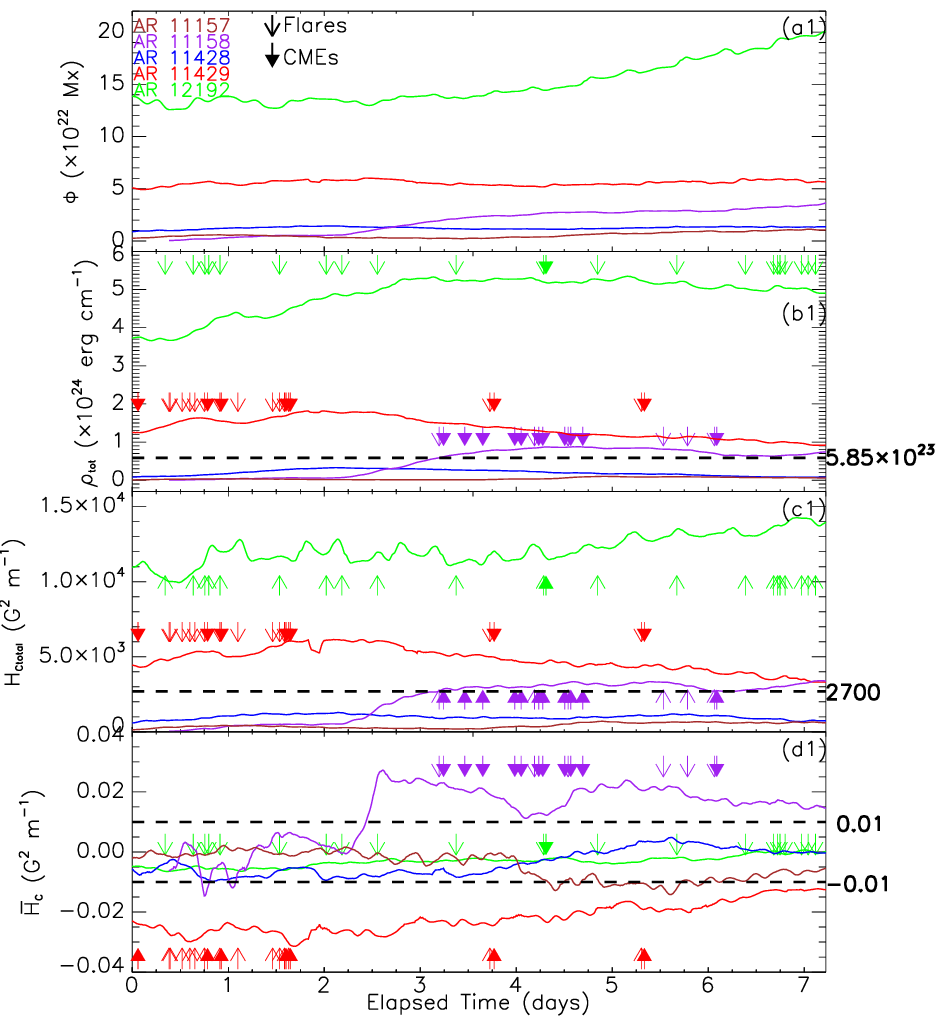}
\caption{ 
Left column: Evolution of the region's photospheric parameters:  
(a) Unsigned magnetic flux $\Phi$, (b) proxy for total photospheric magnetic free energy density $\rho_{total}$, (c) total current helicity ${H_c}_{total}$ ($B_z$ contribution), and (d) mean current helicity $\overline{H_c}$ ($B_z$ contribution). The vertical dashed lines indicate the 14 major events with identifiable source locations,  
while the vertical red line marks the onset time of AR 13668 emergence. 
The horizontal dashed lines indicate the empirical values associated with AR productivity in our sample. 
Right column: 
Same parameters for ARs 11157, 11158, 11428, 11429, and 12192 reproduced from~\citet{Liu_2016}, shown for comparison.  
\copyright~AAS. Reproduced with permission. 
}\label{fig:para} 
\end{center}
\end{figure*}

The flux emergence is reflected by the temporal evolution of the unsigned magnetic flux $\Phi$ of the region (Fig.~\ref{fig:para}a). 
Prior to the emergence of AR 13668, the flux remained relatively stable with a magnitude of $\sim 10^{22}$ Mx. 
After AR 13668 emerged, 
$\Phi$ increased rapidly, 
with the fastest growth occurring between May 7 and May 10, 
during which the most geoeffective CMEs were launched. Moreover, 
$\Phi$ reached a peak value of $1.31\times 10^{23}$ Mx on May 12.

As the flux emerged, 
the photospheric parameters quantifying the region's non-potentiality, 
including $\rho_{total}$, ${H_c}_{total}$, and $\overline{H_c}$, 
all increased dramatically (Fig.~\ref{fig:para}b-d). 
The parameter $\rho_{total}$ grew from a magnitude of $10^{23}$ erg cm$^{-1}$ to a maximum of $3.7\times 10^{24}$ erg cm$^{-1}$ on May 10. 
Similarly, ${H_c}_{total}$ increased from a magnitude of $10^{3}$ G$^2$ m$^{-1}$ to a maximum of $1.5\times 10^{4}$ G$^2$ m$^{-1}$ on May 10, 
while $\overline{H_c}$ rose from a magnitude of $10^{-3}$ G$^2$ m$^{-1}$ to a peak of 0.034 G$^2$ m$^{-1}$ on May 9. 
The fastest growth occurred between May 7 and May 10. 

We also compared the photospheric parameter evolution of the AR complex with that 
in five other ARs, 
NOAA ARs 11157, 11158, 11428, 11429, and 12192 (Fig.~\ref{fig:para}a1-d1), 
which were analyzed by~\citet{Liu_2016} to investigate what determines their flare and CME productivity. 
AR 12192 is well known for producing six confined X-class flares without major CMEs, 
while AR 11158 and 11429 are productive in both flares and CMEs. 
AR 11157 and 11428 are two inert ARs included for comparison.

Although the magnetic flux of AR 13664/8 reached $10^{23}$ Mx on May 9, 
it remained lower than that of AR 12192. 
For the three non-potentiality parameters, 
AR 13664/8 exhibits a large $\rho_{total}$, 
second only to AR 12192, 
a large ${H_c}_{total}$ comparable to that of AR 12192, 
and the highest $\overline{H_c}$ among all ARs during its productive phase. 
These results indicate an exceptionally high non-potentiality of the AR complex, 
even compared with other SARs, 
consistent with the comparison results in~\citet{Kontogiannis_2024a} and ~\citet{Jaswal_2025}.

Moreover, we identified empirical parameter values associated with different levels of AR productivity within our sample. For $\rho_{total}$ and ${H_c}_{total}$, values of approximately $5.85\times 10^{23}$ erg cm$^{-1}$ and $2700$ G$^2$ m$^{-1}$ roughly separate flare-productive ARs 11158, 11429, 12192, and 13664/8 from the two inert ARs. In addition, for $\overline{H_c}$, a value of $0.01$ G$^2$ m$^{-1}$ ($-0.01$ for AR 11429) roughly separates CME-productive ARs 11158, 11429, and 13664/8 from CME-poor ARs 11157, 11428, and 12192. Note that these values are empirical and statistically derived from a limited sample and should not be seen as universal discriminators of AR productivity.

\subsection{Magnetic source locations of major flares and CMEs}\label{subsec:source}

\subsubsection{Magnetic source identification}\label{subsubsec:source}

\begin{figure*}
\begin{center}
\includegraphics[width=1.02\hsize]{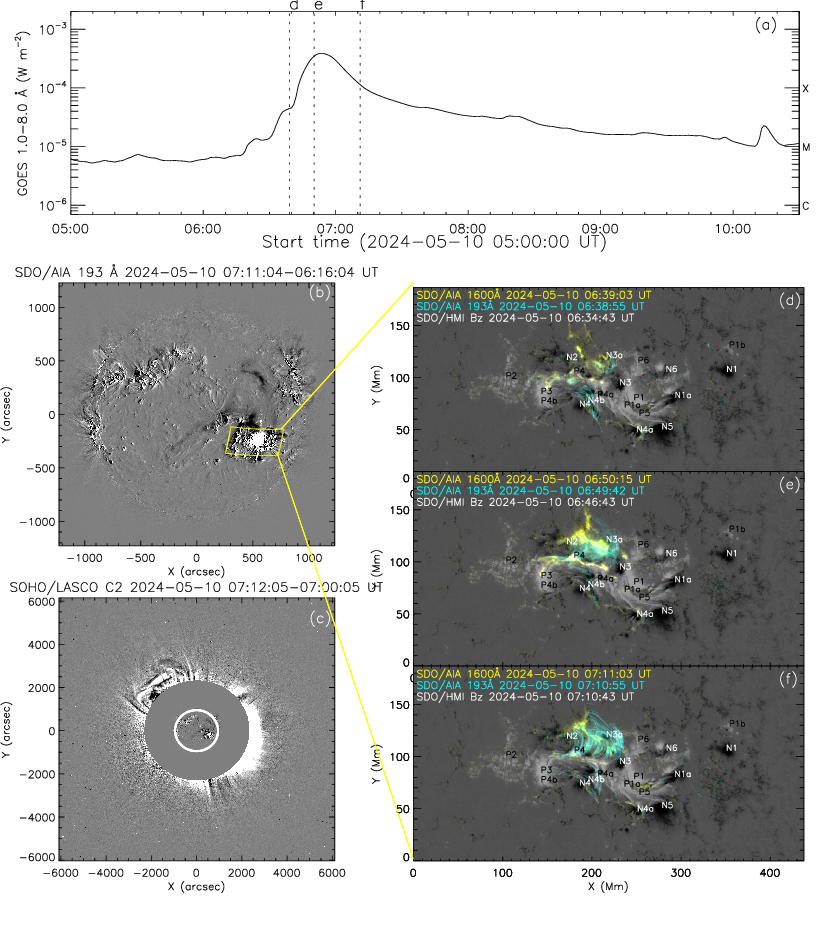}
\caption{
Identification of the X4.0 flare (SOL 2024-05-10 06:27:00 UT; event 10).
(a) GOES soft X-ray flux. 
The vertical dashed lines indicate the timings of panels d--f. 
(b) Full-disk AIA 193 Å image (base-differenced), showing flaring and dimming signatures.  
The yellow box marks the FOV in panels d--f. 
(c) LASCO/C2 coronagraph image showing the associated CME (base-differenced). The white circle marks the solar limb. 
(d)–(f) Composite HMI $B_z$, AIA 1600 Å, and AIA 193 Å maps, 
used to locate the source PIL based on flare ribbons and post-flare loops. 
An animation showing the flare and CME process from  2024-05-10 06:18:04 UT to 2024-05-10 07:47:27 UT is available, 
including panels of GOES soft X-ray fluxes, full-disk AIA 193 Å images, LASCO/C2 coronagraph images, and composite source region maps.  
}\label{fig:source} 
\end{center}
\end{figure*}

We identified the precise source locations of 14 major flares and CMEs (events 1-14 in Table~\ref{tb:eru}) occurring within Stonyhurst longitudes 75$^\circ$E--75$^\circ$W. 
The identification process is illustrated using the X4.0-class flare (SOL 2024-05-10 06:27:00 UT; event 10 in Table~\ref{tb:eru}) as an example (Fig.~\ref{fig:source}). 
First, when the flare appeared in the GOES soft X-ray flux, 
strong flaring activity was observed only in the AR complex (Fig.~\ref{fig:source}a–b), 
confirming its origin in this region. 
Second, the flare was accompanied by clear mass ejection and coronal dimming signatures (Fig.~\ref{fig:source}b). 
Approximately 18 minutes after the flare peak, a partial-halo CME appeared in the LASCO/C2 FOV in the western direction, consistent with the AR on-disk location. Therefore, we associated this CME with the flare. 
Finally, composite images of HMI $B_z$, AIA 1600~\AA, and 193~\AA\ observations reveal flare ribbons on both sides of PIL P3/P4/P4a/P4b–N2/N3a 
and post-flare loops above it (Fig.~\ref{fig:source}d-f). 
We thus identify this PIL as the eruption source, 
hereafter PIL3  (see its detailed evolution in Fig.~\ref{fig:PIL3}).

\begin{figure*}
\begin{center}
\sidecaption
\includegraphics[width=0.70\textwidth]{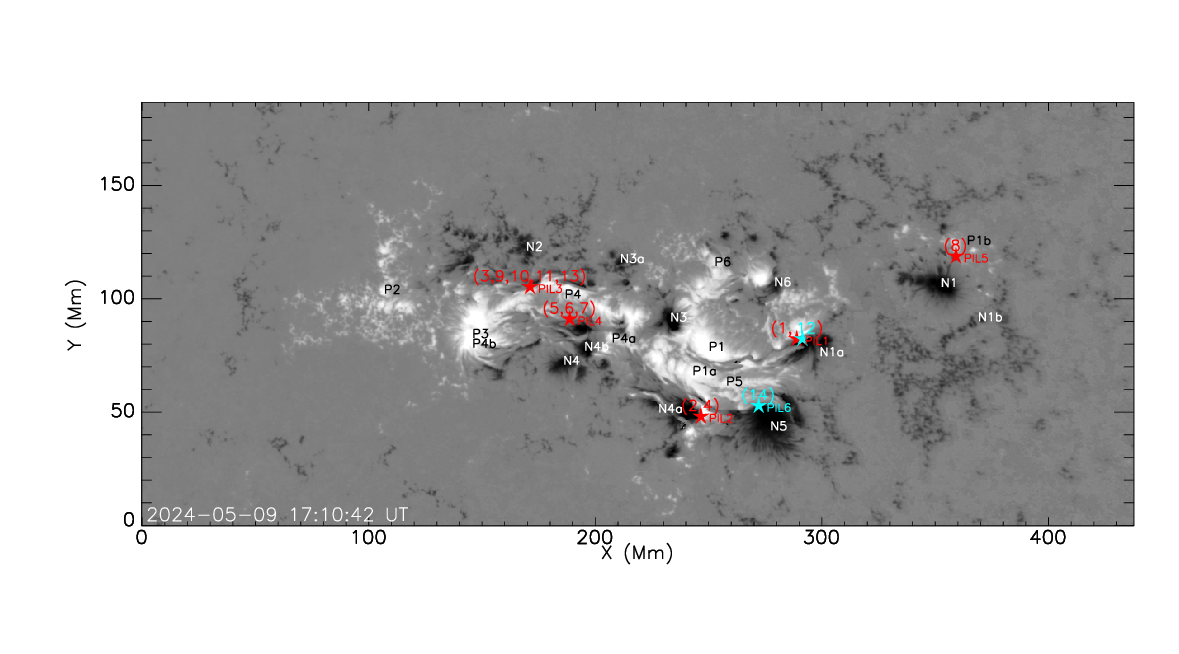}
\caption{Magnetic source locations of 14 major flares and CMEs occurring within Stonyhurst longitudes 75$^\circ$E--75$^\circ$W.  
The asterisks labeling PIL1-PIL6 denote the source cPILs, 
with red and cyan indicating CME and confined-flare sources, respectively. 
Numbers above the asterisks correspond to the event numbers  in Table~\ref{tb:eru}. 
Polarity labels follow the notation in Fig.~\ref{fig:bz}. 
}\label{fig:loc} 
\end{center}
\end{figure*}

The 14 events originated from a total of six PILs (see Fig.~\ref{fig:loc}), a notably large number,  
all of which were aforementioned cPILs: 
PILs P1/P5-N1a, 
P4a/P5-N4/N4a/N4b, P3/P4/P4a/P4b-N2/N3a, P3/P4/P4a/P4b-N4/N4b, P1b-N1, and P5-N4a/N5, 
hereafter PIL1-PIL6. 
Notably, three cPILs produced multiple CMEs: PIL3 generated five CMEs; 
PIL4 produced three; 
and PIL2 generated two. 
The most geoeffective CMEs occurred around May 8. 
Therefore, 
PIL2 and PIL4, which produced the majority of CMEs on May 8, appear to be responsible for these events. 
Note that the event source locations were identified based on flare-related signatures.  
The full eruption of associated CMEs may later involve larger-scale flux systems than the localized sources. 
The only two confined flares (cyan symbols in Fig.~\ref{fig:loc}; events 12 and 14 in Table~\ref{tb:eru}) originated from PIL1 and PIL6.

Interestingly, none of the above eruptions originated from the main interaction interface between AR 13664 and AR 13668, 
i.e., the PIL P1/P6-N3/N3a. 
This may be attributed to the lack of significant shearing between these polarities 
(see Fig.~\ref{fig:bz} associated movie), 
consistent with the collisional shearing scenario~\citep{Chintzoglou_2018}. 
P1/P6-N3/N3a was not the only inter-AR PIL between the two sub-ARs. Another inter-AR PIL segment formed between P5 and N1a, 
which is included as part of the extended cPIL P1/P5-N1a (PIL1; see Fig.~\ref{fig:PIL1}). 
An elongated inter-AR contact interface developed between the same-signed polarities P5 and P1/P1a, 
inhibiting the formation of additional direct inter-AR PILs.

\subsubsection{Evolution of the source PILs}\label{subsubsec:pil}

\begin{figure*}
\begin{center}
\includegraphics[width=1.02\hsize]{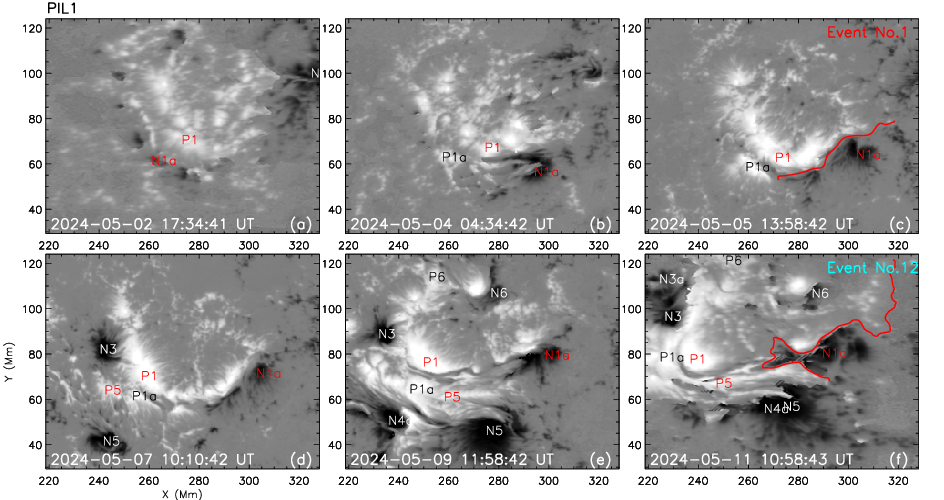} 
\caption{Evolution of the photospheric magnetic field ($B_z$ component) forming PIL1. 
The figure FOV is outlined by a red box labeled ``1'' in Fig.~\ref{fig:bz}f. 
Polarities within the FOV are labeled, 
with those forming the PIL highlighted in red. The red curves in panels (c) and (f) mark the source PILs immediately before events 1 and 12. 
Event numbers are labeled in red for eruptive flares and cyan for confined flares. 
The panels span from initial appearance of the PIL-forming polarities 
(excluding the preexisting N1 and P1) to the last identifiable eruption from the PIL. 
The snapshots are taken from an accompanying online movie spanning from 2024-05-02 17:34:41 UT to 2024-05-11 10:58:43 UT. 
}\label{fig:PIL1} 
\end{center}
\end{figure*}

\begin{figure*}
\begin{center}
\includegraphics[width=1.02\hsize]{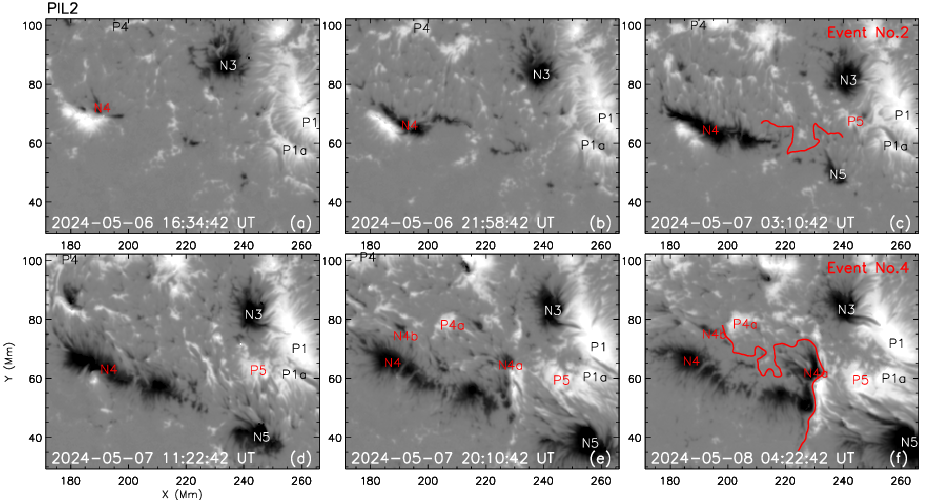}
\caption{Evolution of the photospheric magnetic field forming PIL2. The layout is similar to that of Fig.~\ref{fig:PIL1}. 
The panels are snapshots from an accompanying online movie spanning from 2024-05-06 16:34:42 UT to 2024-05-08 04:22:42 UT. 
}\label{fig:PIL2} 
\end{center}
\end{figure*}

\begin{figure*}
\begin{center}
\includegraphics[width=1.02\hsize]{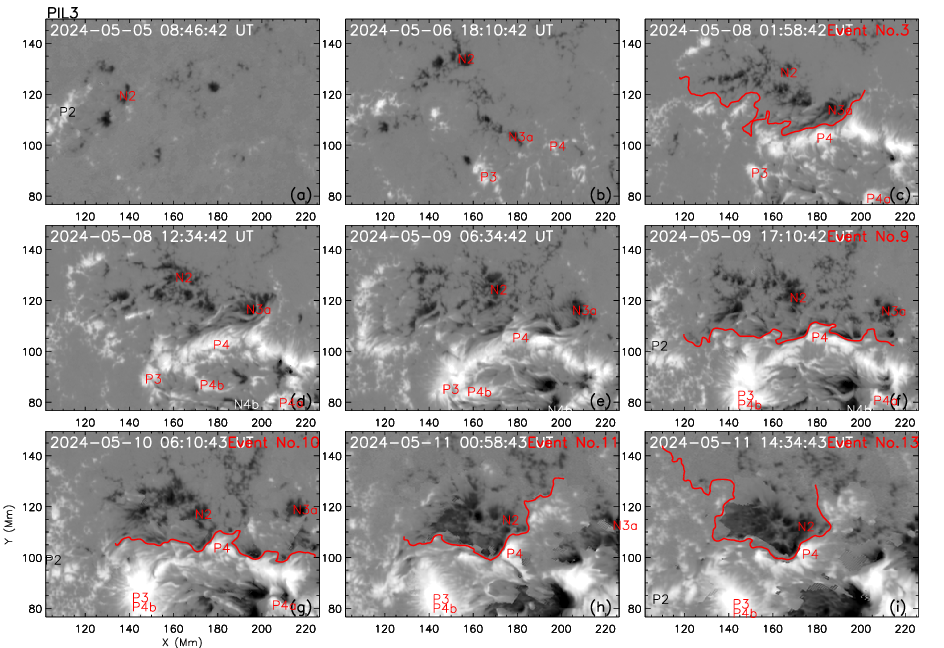}
\caption{Evolution of the photospheric magnetic field ($B_z$ component) 
forming PIL3, with a layout similar to that in Fig.~\ref{fig:PIL1}. 
The panels are snapshots from an accompanying online movie spanning from 2024-05-05 08:46:42 UT to 2024-05-11 14:34:43 UT. 
}\label{fig:PIL3} 
\end{center}
\end{figure*}

\begin{figure*}
\begin{center}
\includegraphics[width=1.02\hsize]{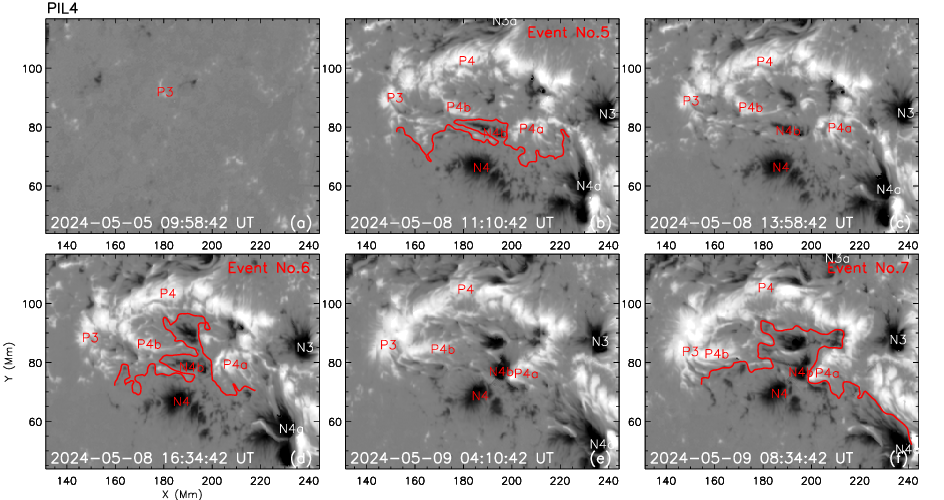}
\caption{Evolution of the photospheric magnetic field ($B_z$ component) 
forming PIL4, with a layout similar to that in Fig.~\ref{fig:PIL1}. 
The panels are snapshots from an accompanying online movie spanning from 2024-05-05 09:58:42 UT to 2024-05-09 08:34:42 UT. 
}\label{fig:PIL4} 
\end{center}
\end{figure*}

\begin{figure*}
\begin{center}
\includegraphics[width=1.02\hsize]{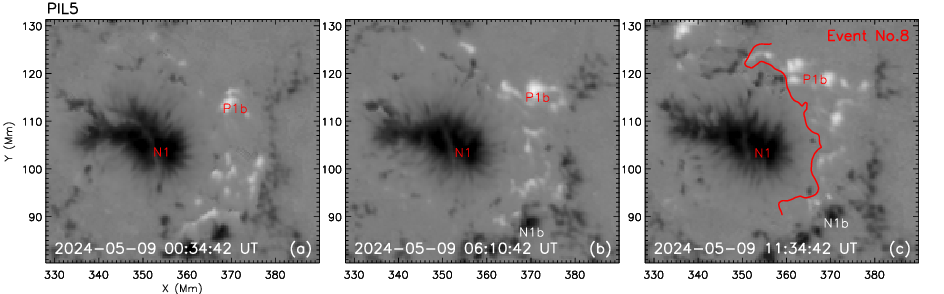}
\caption{Evolution of the photospheric magnetic field ($B_z$ component) forming PIL5, with a layout similar to that in Fig.~\ref{fig:PIL1}. 
The panels are snapshots from an accompanying online movie spanning from 2024-05-09 00:34:42 UT to 2024-05-09 11:34:42 UT. 
}\label{fig:PIL5} 
\end{center}
\end{figure*}

\begin{figure*}
\begin{center}
\includegraphics[width=1.02\hsize]{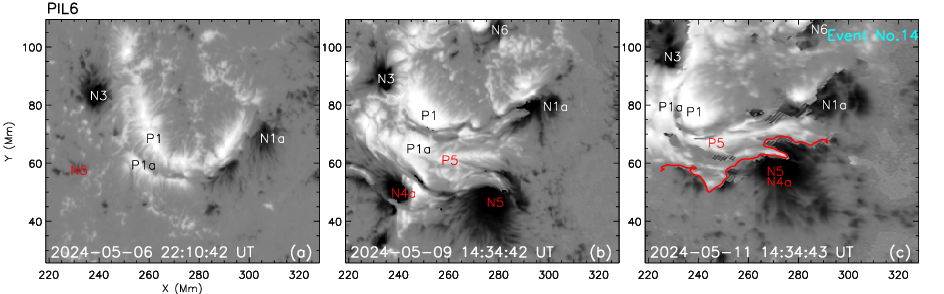} 
\caption{
Evolution of the photospheric magnetic field ($B_z$ component) forming PIL6, with a layout similar to that in Fig.~\ref{fig:PIL1}. 
The panels are snapshots from an accompanying online movie spanning from 2024-05-06 22:10:42 UT to 2024-05-12 15:58:43 UT. 
}\label{fig:PIL6} 
\end{center}
\end{figure*}

We zoomed in on the FOV of each source cPIL to present a clearer view of the magnetic field evolution forming the cPILs. 
The results are shown in Figs.~\ref{fig:PIL1}-\ref{fig:PIL6}. 

PIL1 formed between the nonconjugated polarities P1/P5 and N1a. 
N1a emerged as part of a new bipole near the preexisting P1 (Fig.~\ref{fig:PIL1}a-c), 
leading to direct collision. 
Its separation from conjugate P1a drove sustained shearing against P1 (see movie accompanying Fig.~\ref{fig:PIL1}), 
establishing a classical cPIL. 
After May 9, 
the later-emerged P5 approached P1 and eventually collided with N1a (Fig.~\ref{fig:PIL1}d-f). 
Two identifiable major events originated from this PIL: 
an eruptive C7.5-class flare on May 5, 
about three days after N1a emergence (event 1; Fig.~\ref{fig:PIL1}c) and an X1.5-class confined flare on May 11 (event 12; Fig.~\ref{fig:PIL1}f).

PIL2 exhibits more complex evolution than PIL1 (Fig.~\ref{fig:PIL2} and associated movie). 
Its first major eruption, 
an eruptive C9.7-class flare on May 7 (event 2), 
occurred when the PIL formed between the nonconjugated P5 and N4 regions (Fig.~\ref{fig:PIL2}c). The second major eruption, an X1.0-class eruptive flare on May 8 (event 4),  
occurred after the PIL evolved to involve P4a/P5 and N4/N4a/N4b (Fig.~\ref{fig:PIL2}f). 
Collisional shearing became more pronounced before the second event, especially between P5 and N4a.

PIL3 is a complex cPIL formed between the nonconjugated P3/P4/P4a/P4b and N2/N3a regions (Fig.~\ref{fig:PIL3}). 
It produced five identifiable major eruptions. 
The first was an M3.4-class eruptive flare on May 8 
(event 3; Fig.~\ref{fig:PIL3}c), when P4b had not yet emerged, but collisional shearing was already clear along the PIL.
The next three eruptive events, 
including an X1.1-class flare on May 9 
(event 9; Fig.~\ref{fig:PIL3}f), an X4.0-class flare on May 10 
(event 10; Fig.~\ref{fig:PIL3}g), and an X5.8-class flare on May 11 (event 11; Fig.~\ref{fig:PIL3}h) occurred successively with similar magnitude and morphology, 
making them homologous events. 
A final M8.9-class eruptive flare followed on May 11 (event 13; Fig.~\ref{fig:PIL3}i).
During this period, P4b emerged into the system, while N3a gradually moved away after May 10. 
Nonetheless, strong collisional shearing persisted 
throughout these eruptions.

PIL4 is another complex cPIL, formed between P3/P4/P4a/P4b and N4/N4b (Fig.~\ref{fig:PIL4}). 
It produced three major eruptions:  
an M8.7-class eruptive flare on May 8 
(event 5; Fig.~\ref{fig:PIL4}b), 
an X1.0-class eruptive flare later the same day (event 6; Fig.~\ref{fig:PIL4}d), and an X2.3-class eruptive flare on  May 9 (event 7; Fig.~\ref{fig:PIL4}f). 
The three eruptions are homologous events occurring successively with similar magnitude and morphology as well. 
The strongest collisional shearing developed mainly between P4a and N4b (see movie accompanying Fig.~\ref{fig:PIL4}).

Both PIL5 and PIL6 produced a single major event. 
The collisional PIL5 formed between the nonconjugated P1b and N1 (Fig.~\ref{fig:PIL5}) 
and produced an M3.1-class eruptive flare on May 9 
(event 8; Fig.~\ref{fig:PIL5}c). 
PIL6 developed between P5 and N4a/N5, producing an X1.0-class confined flare (event 14; Fig.~\ref{fig:PIL6}c). 
In PIL6, collisional shearing occurred mainly between the nonconjugated P5 and N4a regions, confirming its cPIL nature.

For PIL6, although its temporal evolution and role as the source of event 14 can be established, 
the vector magnetic field data became relatively noisy at the time of event 14 (2024-05-12 16:11:00 UT), 
when the region was close to the solar limb. 
We therefore used the vector magnetogram before the previous event (at 2024-05-11 14:34:43 UT)  
to identify PIL6 (Fig.~\ref{fig:PIL6}c) and calculate the decay index. 
The large-scale magnetic configuration (e.g., the potential field used in the decay index calculation) is not expected to change significantly over this timescale ($\sim1$ day).

\subsubsection{Decay index}\label{subsubsec:decay}
\begin{figure*}
\begin{center}
\includegraphics[width=1.02\hsize]{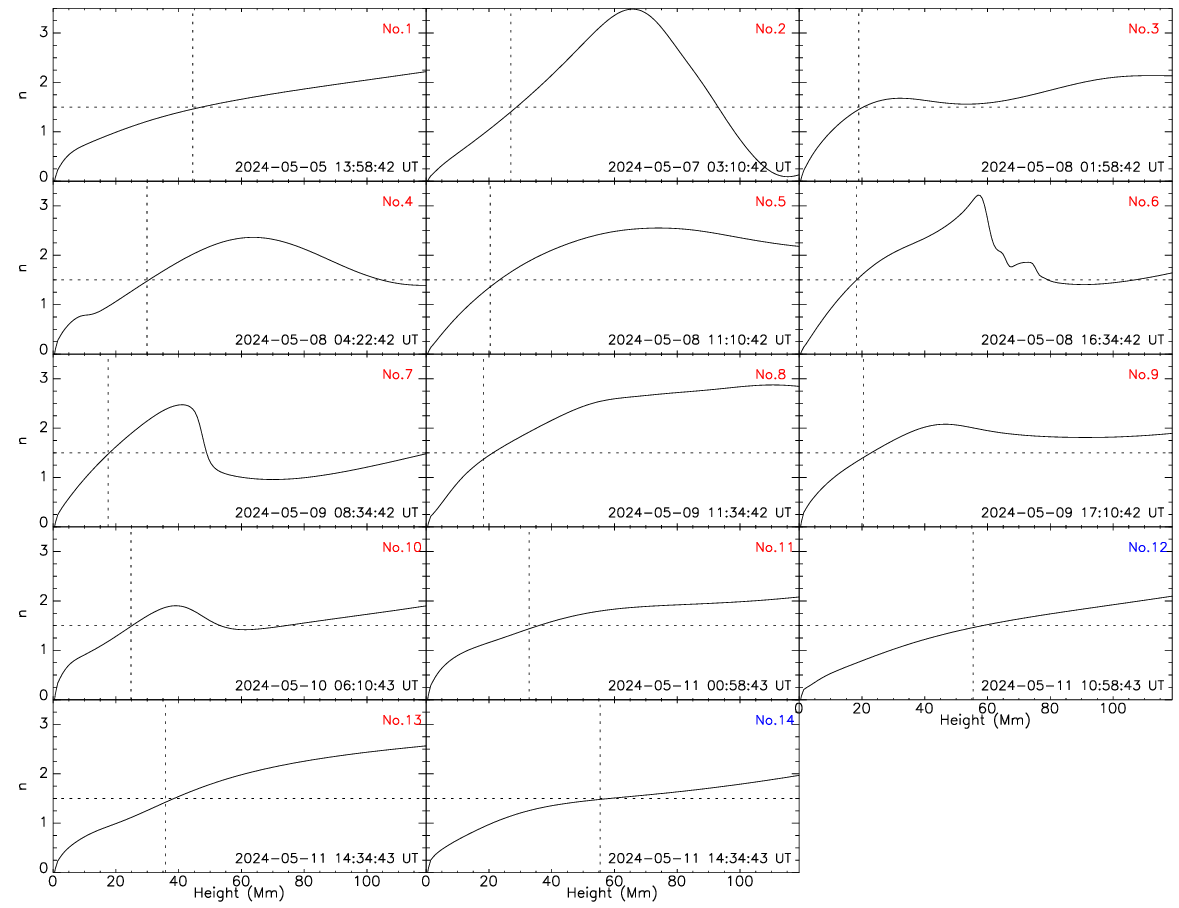} 
\caption{
Decay index distribution above the source PILs (shown in Figs.~\ref{fig:PIL1}-\ref{fig:PIL6}) right before the onset of the 14 major flares and CMEs with identifiable magnetic sources. 
The numbers correspond to the event numbers in Table~\ref{tb:eru}. 
Red represents the CMEs, and cyan represents the confined flares. 
The x-axis indicates height above the PIL, 
while the y-axis shows the decay index, $n$. 
Each data point represents PIL-averaged $n$ at a given height. 
The horizontal dashed lines mark the torus instability threshold $n=1.5$, 
while the vertical dashed lines indicate the critical heights where $n$ first reaches $1.5$. }\label{fig:decay}  
\end{center}
\end{figure*}

The decay index ($n$) distributions above the source PILs (shown in Figs.~\ref{fig:PIL1}-\ref{fig:PIL6}) prior to the 14 events are shown in Fig.~\ref{fig:decay}. 
The critical heights, where $n$ reaches 1.5, range from 18 Mm to 56 Mm. 
Although the sample includes only two confined flares, 
a clear tendency emerges: CME source PILs have lower critical heights. 
Specifically, the critical heights above the CME source PILs  are all below 45 Mm,  
with a median of 25 Mm, whereas those for the confined flares are all higher than 55 Mm.

\subsection{Waiting time distributions}\label{subsec:wt}

\begin{figure}
\begin{center}
\includegraphics[width=\columnwidth, trim=22 2 12 0, clip]{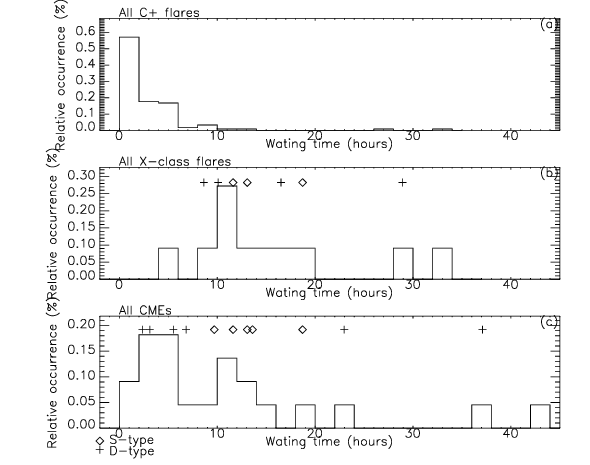}
\caption{
Waiting time distribution for 
(a) all 120 flares above C-class, 
(b) all 12 X-class flares, and (c) all 23 CMEs. 
The x-axis represents the waiting times, 
while the y-axis shows the relative occurrence within each subsample. 
In panels (b)–(c), the diamonds denote 
the waiting times of identifiable S-type events 
(from the same cPIL), 
while the pluses denote the identifiable D-type events 
(from different cPILs).}\label{fig:wtime} 
\end{center}
\end{figure}

Waiting time distributions are shown in Fig.~\ref{fig:wtime}. 
For all 120 flares above C-class, 
the distribution roughly follows an exponential trend (Fig.~\ref{fig:wtime}a), 
with a median waiting time of 1.5 hours, 
suggesting that their occurrence is broadly consistent with a Poisson-like process, 
i.e., independent and random occurrence.

For all 12 X-class flares, 
the distribution differs markedly from an exponential trend (Fig.~\ref{fig:wtime}b). 
It peaks around 11 hours, with a median of 13 hours,  
which may indicate some physical linkage among these large events. 
The small sample size prevents us from assigning a specific functional form or trend to the distribution.

For all 23 CMEs, the waiting time distribution exhibits two peaks at 4 hours and 11 hours (Fig.~\ref{fig:wtime}c), with a median of 10 hours. 
A similar double-peaked distribution is reported by~\citet{Liu_2017} for a much larger sample of 142 homologous and quasi-homologous CMEs, 
with peaks at 1.5 and 7.5 hours. 
Their analysis associates the shorter peak with D-type CMEs (from different sources), which are triggered by disturbances from nearby eruptions~\citep{Torok_2011}, and the longer peak with S-type CMEs (from the same source), which involve recurrent replenishment and release of the magnetic free energy.  
Although our dataset is 
much smaller and limited to a single region, 
the two peaks may still suggest similar underlying processes, 
with the 4- and 11-hour peaks likely corresponding to D-type and S-type events. 
The difference in peak values between the two studies 
may mainly reflect the different sample sizes.

We further examined the waiting times of the 14 major events with identifiable sources and marked them in Fig.~\ref{fig:wtime} 
(diamonds for S-type and plus symbols for D-type). 
Among the 12 CMEs, PIL3 produced four, and PIL4 produced three successive CMEs, yielding five S-type waiting times. 
Four (ranging from 9.7 to 13.6 hours) cluster near the 11-hour peak, which we tentatively associate with S-type events. 
Similarly, among the six identifiable D-type waiting times, four (ranging from 2.4 to 6.8 hours) cluster around the 4-hour peak, which we associate with D-type CMEs.  
These results support the tentative association of the two peaks with S-type and D-type CMEs. 
In contrast, the waiting times of X-class flares show no clear clustering.

\section{Discussion}\label{sec:sum}

In this study, we investigated why the AR complex 13664/8 is so productive, particularly in CMEs. 
We examined the photospheric magnetic evolution, 
eruption sources, and event waiting times. 
Our main results are as follows: 

\begin{itemize}

\item[1)] 
The AR complex initially appeared as an overall bipolar region containing only AR 13664, 
producing no large event before May 4. 
An increase in its productivity followed the emergence of AR 13668 on May 4, after which 12 X-class flares and 23 CMEs occurred within 12 days. 
The rapid and complex flux emergence increased the region's magnetic flux and complexity, resulting in six cPILs formed through collision and shearing between nonconjugated polarities.

\item[2)]
Photospheric magnetic parameters show that the region's non-potentiality increased dramatically during the rapid flux emergence. 
Comparison with five other ARs reveals that both flare-rich and CME-rich ARs exhibit higher overall non-potentiality, 
characterized by large total magnetic free energy densities 
and total current helicities, 
while CME-rich ARs exhibit larger mean current helicities.

\item[3)]  
The 14 major events with identifiable source locations originated from six PILs, all of which are cPILs. 
The decay index distribution reveals lower critical heights for torus instability above CME sources, 
indicating more rapid background decay above eruptive flares.

\item[4)] 
The waiting time distribution for all flares is exponential-like,  
while that for the 12 X-class flares shows a single peak at 11 hours. 
In contrast, 
all 23 CMEs exhibit two distinct peaks at 4 and 11 hours. 
Among the 12 source-identifiable CMEs, 
D-type events (from different sources) tend to cluster around the shorter peak, 
while S-type events (from the same source) tend to cluster around the longer peak. 
The X-class flare peak coincides with the longer CME peak. 

\end{itemize}

\subsection{High non-potentiality and rapidly decaying overlying field}\label{subsec:dis-non-potentiality}

The region size is almost comparable to that of the flare-rich but CME-poor AR 12192,  
a ``big but mild'' AR, with large total non-potentiality but weaker core-region non-potentiality, and stronger overlying confinement.    
These are reflected in large extensive but small intensive non-potentiality parameters, 
along with lower decay index~\citep{Sun_2015, Liu_2016}. 
Extensive parameters generally scale with AR size, 
whereas intensive ones do not, 
and CME productivity has been found to correlate more strongly with intensive non-potentiality parameters~\citep{Welsch_2009, Bobra_Ilonidis_2016}.

The properties of AR 13664/8 are consistent with the above picture. 
It exhibits not only large extensive parameters 
but also a large intensive parameter, 
indicating strong non-potentiality on both global and localized scales. 
The large total non-potentiality provides a necessary condition for both flare and CME production, while the high mean current helicity further indicates the presence of mature, strongly sheared or twisted core field that may serve as CME seeds~\citep{Liu_2016}. 
In our sample of six ARs, the mean photospheric current helicity shows the clearest association with CME productivity 
and may even differentiate between CME-rich and CME-poor phases within the same AR. 
The robustness of the correlation remains to be tested in a larger sample.  

A similar comparative study conducted by~\citet{Kontogiannis_2024a} finds that AR 13664/8 exhibits the largest non-neutralized currents among several highly productive ARs, including ARs 12192, 11158, and 12673. 
This suggests that the AR complex departs from a force-free state, 
which may favor CME occurrence, 
providing a complementary perspective to our results.

In addition, the rapid decay of overlying fields, reflected by the lower critical heights for torus instability above CME sources, serves as another necessary condition for CMEs. 
In our sample, 
a critical height of 45 Mm empirically separates CMEs from confined flares, 
consistent with previous findings~\citep[e.g.,][]{Wangd_2017, Baumgartner_2018}.  

In summary, AR 13664/8 has sufficient non-potentiality and a rapidly decaying constraining field. The two are necessary conditions for CMEs, as established in the SDO era.

\subsection{High magnetic complexity}\label{subsec:dis-complexity}
\subsubsection{Multiple cPILs}\label{subsubsec:dis-complexity-cpil}

In addition, 
AR 13664/8 exhibits unusually high 
magnetic complexity, with dynamical evolution dominated by 
collisional shearing occurring at multiple locations. 
This complexity, manifested by multiple interacting bipoles, 
exceeds that of many previously reported CME-productive ARs such as AR 11158 and AR 11429~\citep{Liu_2016}. 
The emergence of multiple bipoles produced numerous cPILs, 
as conjugated polarities naturally separated and allowed nonconjugated opposite-signed polarities to collide, shear, and cancel, thereby driving flares and CMEs~\citep{Chintzoglou_2018}. 
Here, all 14 major events with identifiable sources originated from as many as six cPILs, 
supporting the above scenario.

However, 
collision alone did not guarantee major eruptions. 
For example, at the noneruptive inter-AR PIL P1/P6-N3/N3a, 
significant shearing was absent: 
the two ARs appeared locked here after collision, 
with little relative motion between the nonconjugated polarities to sustain shearing after their full emergence. 
This further underscores the critical role of shearing.

P1/P6-N3/N3a is not the only interface between AR 13664 and AR 13668. 
Another PIL formed as part of the cPIL P1/P5-N1a,  
which produced two large events, 
consistent with the inter-AR interaction scenario~\citep{Toriumi_2016, Toriumi_2017}. The remaining sources were not inter-AR PILs, 
but their formation all involved bipole interactions, as clarified above and in previous studies~\citep{Sun_2024, Kontogiannis_2024a, Jaswal_2025}. 
These highlight that sustained shearing---regardless of whether the PIL is inter-AR---may be a key factor driving eruptions.

This study focuses on the macroscopic magnetic configuration governing the CME productivity of the AR complex. While the presence and overall role of collisional shearing are established, 
detailed properties of individual cPILs---such as their collision and shearing 
evolution, relation to eruption timing, and possible correlation with flare magnitude---remain beyond the scope of this work 
and warrant further investigation.

\subsubsection{Implication of waiting time distributions}\label{subsubsec:dis-complexity-wt}

Such high complexity with multiple dynamically evolving cPILs 
may enhance the CME productivity in two ways. 
First, multiple cPILs increase the number of sites for recurrent CMEs. 
Second, a CME from one site may disturb nearby flux systems, 
triggering additional CMEs~\citep{Torok_2011, Zuccarello_2014}. 
Waiting time distributions provide supporting evidence.

For all flares above C-class, 
the waiting time distribution is approximately exponential, 
suggesting they occurred largely independently and randomly. 
In contrast, 
power-law distributions have been reported by~\citet{Wheatland_2000} and~\citet{Boffetta_1999}, but with different interpretations. 
\citet{Wheatland_2000} analyzed 25 years of full-disk flares 
and attributes the power law to a time-dependent Poisson process. 
In this scenario, an exponential distribution is expected for short observation periods, such as for a single AR, since the flare rates lack sufficient variation to produce a power law. 
\citet{Boffetta_1999}, however, attributes the power-law distribution of successive same-source flares to long-term correlations, possibly related to magnetohydrodynamic (MHD) turbulence. 
Our result is consistent with~\citet{Wheatland_2000}, 
indicating that for this region, 
flares of all classes are largely independent events.

In contrast, 
the peaks for large flares and CMEs indicate intrinsic physical correlations among major events. 
The clustering of identifiable S-type CMEs around the longer CME peak supports recurrent free energy buildup and release at multiple cPILs on a similar timescale,  
while clustering of identifiable D-type CMEs around the shorter peak suggests CME triggering by nearby CME disturbances 
also on a similar timescale.

This interpretation requires that CME waiting times are not simply inherited from the associated flares with varying magnitudes. 
While large flares require a certain time for energy replenishment, 
weaker flares occur more frequently (Fig.~\ref{fig:wtime}a). 
The 11-hour peak in the X-class flare distribution coincides with the longer CME peak, 
reflecting an energy replenishment timescale for large events.  
For the shorter CME peak (4 hours), 
most associated flares are weaker. 
If this peak were mainly controlled by 
flare waiting times, 
it would be expected to occur close to the typical waiting time for all flares ($\sim$1.5 hours), 
but it is significantly longer. 
Combined with the clustering of identifiable S- and D-type CMEs around the two peaks, 
these support that the double-peaked CME distribution more likely reflects two 
CME mechanisms (energy replenishment and nearby disturbance), 
rather than associated flare waiting times. 
Note that this interpretation remains tentative due to the limited sample size.

Although magnetic non-potentiality and complexity are closely related, they do not emphasize the same aspects of the AR complex here. 
The total non-potentiality parameters (Sect.~\ref{subsec:dis-non-potentiality}) quantify the overall amount of non-potentiality, 
whereas magnetic complexity describes how this non-potentiality is spatially organized and dynamically supplied. 
Multiple cPILs, together with the different peaks of CME waiting times, 
indicate that free energy was not concentrated at a single location 
but was distributed and supplied through sustained collision and shearing at multiple locations, thereby enhancing the CME productivity. 
Among the examined magnetic parameters, 
mean current helicity may be more closely related to the localized complexity, 
as it reflects the relative, rather than the total non-potentiality.

\subsection{Toward a physically motivated definition of magnetic complexity}\label{subsec:dis-complexity-definition}

A more precise definition of magnetic complexity is still needed. 
In AR 13664/8,  
the high complexity arises from the unusually large number of dynamically interacting bipoles, which produced multiple source cPILs. 
However, several widely used classification schemes do not fully capture these properties.

The Hale classification, 
which characterizes AR magnetic complexity~\citep{Hale_1919, Kunzel_1960}, distinguishes sunspot configuration  
but does not specify 
how many bipoles are involved and 
how many interaction interfaces (e.g., cPILs) are formed. 
The more sophisticated McIntosh classification~\citep{McIntosh_1990}, developed for flare prediction,  
considers sunspot Zurich class, penumbra development of the principal spot, and compactness of the sunspot group. However, it 
does not explicitly describe the dynamical bipole interactions. 
\citet{McCloskey_2018} further incorporated the temporal evolution of sunspot groups---the transitions between  McIntosh classes---and showed that upward transitions significantly improve C-class and M-class flare forecasting. 
Nevertheless, this approach also does not characterize bipole interactions in a physical sense.

A more precise definition of complexity---one that incorporates not only the overall level of non-potentiality but also its spatial concentrations (e.g., the number of cPILs) 
and its dynamical evolution (e.g., magnitude of collisional shearing)---is essential. Such a definition may enable more accurate predictions of AR productivity.

\section{Summary and conclusions}\label{sec:conc} 

In this study, we investigated why AR 13664/8 was exceptionally productive, particularly in producing CMEs. We analyzed the photospheric magnetic evolution of the AR complex, the source locations of its major flares and CMEs, the decay of the overlying magnetic field above the source PILs, and the waiting time distributions of flares and CMEs. The results show that, in addition to sufficient non-potentiality and rapidly decaying overlying magnetic fields, the unusually high magnetic complexity of the region, manifested by dynamical collisional shearing at multiple cPILs, played a critical role in sustaining its extreme CME productivity.

\begin{acknowledgements}
We thank our anonymous referee for the constructive comments that significantly improved the manuscript. 
We acknowledge the {\it SDO}, {\it SOHO}, and {\it GOES} missions for providing quality observations. 
We also acknowledge the support from National Space Science Data Center, National Science \& Technology Infrastructure of China (www.nssdc.ac.cn). 
This work is supported by the National Natural Science Foundation of China (Grant Nos. 42188101, 12273123, 42174213,  42474204, 42521007, 42474221, and 42441808), the Guangdong Basic and Applied Basic Research Foundation (2023A1515030185), and the Strategic Priority Research Program of the Chinese Academy of Sciences (Grant No. XDB0560000). 

\end{acknowledgements}

\bibliographystyle{aa}
\bibliography{AR13664}

@article{Torok_2011,
	title = {A {MODEL} {FOR} {MAGNETICALLY} {COUPLED} {SYMPATHETIC} {ERUPTIONS}},
	volume = {739},
	issn = {2041-8205},
	doi = {10.1088/2041-8205/739/2/L63},
	number = {2},
	journal = {ApJL},
	author = {Török, T. and Panasenco, O. and Titov, V. S. and Mikić, Z. and Reeves, K. K. and Velli, M. and Linker, J. A. and De Toma, G.},
	month = oct,
	year = {2011},
	pages = {L63},
}

@article{Toriumi_2019,
	title = {Flare-productive active regions},
	volume = {16},
	issn = {2367-3648},
	url = {https://doi.org/10.1007/s41116-019-0019-7},
	doi = {10.1007/s41116-019-0019-7},
	number = {1},
	journal = {LRSP},
	publisher = {Springer International Publishing},
	author = {Toriumi, Shin and Wang, Haimin},
	month = dec,
	year = {2019},
	pages = {3},
}

@article{Pesnell_2012,
	title = {The {Solar} {Dynamics} {Observatory} ({SDO})},
	volume = {275},
	issn = {0038-0938},
	url = {http://link.springer.com/10.1007/s11207-011-9841-3},
	doi = {10.1007/s11207-011-9841-3},
	number = {1-2},
	journal = {SoPh},
	author = {Pesnell, W.{\textasciitilde}D. Dean and Thompson, B.{\textasciitilde}J. J. and Chamberlin, P.{\textasciitilde}C. C.},
	month = jan,
	year = {2012},
	pages = {3--15},
}

@article{Liu_2021,
	title = {The {Source} {Locations} of {Major} {Flares} and {CMEs} in {Emerging} {Active} {Regions}},
	volume = {909},
	copyright = {All rights reserved},
	issn = {0004-637X},
	url = {https://iopscience.iop.org/article/10.3847/1538-4357/abde37},
	doi = {10.3847/1538-4357/abde37},
	number = {2},
	journal = {ApJ},
	author = {Liu, Lijuan and Wang, Yuming and Zhou, Zhenjun and Cui, Jun},
	month = mar,
	year = {2021},
	pages = {142},
}

@article{Gan_2019,
	title = {Advanced {Space}-based {Solar} {Observatory} ({ASO}-{S}): an overview},
	volume = {19},
	issn = {1674-4527},
	url = {http://iopscience.iop.org/1674-4527/14/4/006},
	doi = {10.1088/1674-4527/19/11/156},
	language = {en-US},
	number = {11},
	journal = {RAA},
	author = {Gan, Wei-Qun and Zhu, Cheng and Deng, Yuan-Yong and Li, Hui and Su, Yang and Zhang, Hai-Ying and Chen, Bo and Zhang, Zhe and Wu, Jian and Deng, Lei and Huang, Yu and Yang, Jian-Feng and Cui, Ji-Jun and Chang, Jin and Wang, Chi and Wu, Ji and Yin, Zeng-Shan and Chen, Wen and Fang, Cheng and Yan, Yi-Hua and Lin, Jun and Xiong, Wei-Ming and Chen, Bin and Bao, Hai-Chao and Cao, Cai-Xia and Bai, Yan-Ping and Wang, Tao and Chen, Bing-Long and Li, Xin-Yu and Zhang, Ye and Feng, Li and Su, Jiang-Tao and Li, Ying and Chen, Wei and Li, You-Ping and Su, Ying-Na and Wu, Hai-Yan and Gu, Mei and Huang, Lei and Tang, Xue-Jun},
	month = nov,
	year = {2019},
	pages = {156},
}

@article{Domingo_2000,
	title = {The scientific payload of the space-based {Solar} and {Heliospheric} {Observatory} ({SOHO})},
	volume = {70},
	issn = {0038-6308},
	url = {http://link.springer.com/10.1007/BF00777835},
	doi = {10.1007/BF00777835},
	number = {1-2},
	journal = {SSRv},
	author = {Domingo, V and Fleck, B and Poland, A. I.},
	month = oct,
	year = {1994},
	pages = {7--12},
}

@article{Vemareddy_2019,
	title = {Very {Fast} {Helicity} {Injection} {Leading} to {Critically} {Stable} {State} and {Large} {Eruptive} {Activity} in {Solar} {Active} {Region} {NOAA} 12673},
	volume = {872},
	issn = {1538-4357},
	url = {http://arxiv.org/abs/1901.09358},
	doi = {10.3847/1538-4357/ab0200},
	number = {2},
	journal = {ApJ},
	author = {Vemareddy, P.},
	month = feb,
	year = {2019},
	pages = {182},
}

@article{Sun_2015,
	title = {Why {Is} the {Great} {Solar} {Active} {Region} 12192 {Flare}-{Rich} {But} {CME}-{Poor}?},
	volume = {804},
	url = {http://arxiv.org/abs/1502.06950},
	doi = {10.1088/2041-8205/804/2/L28},
	journal = {ApJL},
	author = {Sun, Xudong and Bobra, Monica G. and Hoeksema, J. Todd and Liu, Yang and Li, Yan and Shen, Chenglong and Couvidat, Sebastien and Norton, Aimee A. and Fisher, George H.},
	month = feb,
	year = {2015},
	pages = {l28},
}

@article{Scherrer_2012,
	title = {The {Helioseismic} and {Magnetic} {Imager} ({HMI}) {Investigation} for the {Solar} {Dynamics} {Observatory} ({SDO})},
	volume = {275},
	issn = {0038-0938},
	url = {http://link.springer.com/10.1007/978-1-4614-3673-7_10},
	doi = {10.1007/s11207-011-9834-2},
	number = {1-2},
	journal = {SoPh},
	publisher = {Springer US},
	author = {Scherrer, P. H. and Schou, J. and Bush, R. I. and Kosovichev, A. G. and Bogart, R. S. and Hoeksema, J. T. and Liu, Y. and Duvall, T. L. and Zhao, J. and Title, A. M. and Schrijver, C. J. and Tarbell, T. D. and Tomczyk, S.},
	month = jan,
	year = {2012},
	pages = {207--227},
}

@article{Van_2015,
	title = {Evolution of {Active} {Regions}},
	volume = {12},
	issn = {2367-3648},
	url = {http://link.springer.com/10.1007/lrsp-2015-1},
	doi = {10.1007/lrsp-2015-1},
	number = {1},
	journal = {LRSP},
	author = {van Driel-Gesztelyi, Lidia and Green, Lucie May},
	month = dec,
	year = {2015},
	pages = {1},
}

@article{Zhang_Wang_2002,
	title = {Are {Homologous} {Flare}–{Coronal} {Mass} {Ejection} {Events} {Triggered} by {Moving} {Magnetic} {Features}?},
	volume = {566},
	issn = {0004637X},
	url = {http://adsabs.harvard.edu/abs/2002ApJ...566L.117Z},
	doi = {10.1086/339660},
	number = {2},
	journal = {ApJ},
	author = {Zhang, Jun and Wang, Jingxiu},
	month = feb,
	year = {2002},
	pages = {L117--L120},
}

@article{Webb_2012,
	title = {Coronal {Mass} {Ejections}: {Observations}},
	volume = {9},
	issn = {1614-4961},
	url = {http://link.springer.com/10.12942/lrsp-2012-3},
	doi = {10.12942/lrsp-2012-3},
	number = {1},
	journal = {LRSP},
	author = {Webb, David F. and Howard, Timothy A.},
	year = {2012},
	pages = {3},
}

@article{Chengx_2011a,
	title = {A {COMPARATIVE} {STUDY} {OF} {CONFINED} {AND} {ERUPTIVE} {FLARES} {IN} {NOAA} {AR} 10720},
	volume = {732},
	issn = {0004-637X},
	url = {http://stacks.iop.org/0004-637X/732/i=2/a=87?key=crossref.f6ad1c2320ef5af99b5971b72dda9e0b},
	doi = {10.1088/0004-637X/732/2/87},
	number = {2},
	journal = {ApJ},
	author = {Cheng, X. and Zhang, J. and Ding, M. D. and Guo, Y. and Su, J. T.},
	month = may,
	year = {2011},
	pages = {87},
}

@article{Guo_2010,
	title = {{DRIVING} {MECHANISM} {AND} {ONSET} {CONDITION} {OF} {A} {CONFINED} {ERUPTION}},
	volume = {725},
	issn = {2041-8205},
	url = {http://stacks.iop.org/2041-8205/725/i=1/a=L38?key=crossref.000d80c9b93c82b4ace49a4e454170e6},
	doi = {10.1088/2041-8205/725/1/L38},
	number = {1},
	journal = {ApJL},
	author = {Guo, Y. and Ding, M. D. and Schmieder, B. and Li, H. and Török, T. and Wiegelmann, T.},
	month = dec,
	year = {2010},
	pages = {L38--L42},
}

@article{Falconer_etal_2006,
	title = {Magnetic {Causes} of {Solar} {Coronal} {Mass} {Ejections}: {Dominance} of the {Free} {Magnetic} {Energy} over the {Magnetic} {Twist} {Alone}},
	volume = {644},
	issn = {0004-637X},
	url = {http://stacks.iop.org/0004-637X/644/i=2/a=1258},
	doi = {10.1086/503699},
	number = {2},
	journal = {ApJ},
	author = {Falconer, D. A. and Moore, R. L. and Gary, G. a.},
	month = jun,
	year = {2006},
	pages = {1258--1272},
}

@article{Chen_Wang_2011,
	title = {Statistical study of coronal mass ejection source locations: 2. {Role} of active regions in {CME} production},
	volume = {116},
	issn = {01480227},
	url = {http://doi.wiley.com/10.1029/2011JA016844},
	doi = {10.1029/2011JA016844},
	number = {A12},
	journal = {JGRA},
	author = {Chen, Caixia and Wang, Yuming and Shen, Chenglong and Ye, Pinzhong and Zhang, Jie and Wang, S.},
	month = dec,
	year = {2011},
	pages = {n/a--n/a},
}

@article{Wang_2013b,
	title = {Waiting {Times} of {Quasi}-{Homologous} {Coronal} {Mass} {Ejections} {From} {Super} {Active} {Regions}},
	volume = {763},
	copyright = {All rights reserved},
	issn = {2041-8205},
	url = {http://stacks.iop.org/2041-8205/763/i=2/a=L43?key=crossref.c892bc2e1c3080e243f618e18678f87f},
	doi = {10.1088/2041-8205/763/2/L43},
	number = {2},
	journal = {ApJL},
	author = {Wang, Yuming and Liu, Lijuan and Shen, Chenglong and Liu, Rui and Ye, Pinzhong and Wang, S.},
	month = feb,
	year = {2013},
	pages = {L43},
}

@article{Leka_Barnes_2007,
	title = {Photospheric {Magnetic} {Field} {Properties} of {Flaring} versus {Flare}‐quiet {Active} {Regions}. {IV}. {A} {Statistically} {Significant} {Sample}},
	volume = {656},
	issn = {0004-637X},
	url = {http://stacks.iop.org/0004-637X/656/i=2/a=1173},
	doi = {10.1086/510282},
	number = {2},
	journal = {ApJ},
	author = {Leka, K. D. and Barnes, G.},
	month = feb,
	year = {2007},
	pages = {1173--1186},
}

@article{Liu_2016,
	title = {Why {Is} a {Flare}-{Rich} {Active} {Region} {Cme}-{Poor}?},
	volume = {826},
	copyright = {All rights reserved},
	issn = {1538-4357},
	url = {http://stacks.iop.org/0004-637X/826/i=2/a=119?key=crossref.6706f3ae093e84ab22d06614d6c839ca},
	doi = {10.3847/0004-637X/826/2/119},
	language = {en-US},
	number = {2},
	journal = {ApJ},
	publisher = {IOP Publishing},
	author = {Liu, Lijuan and Wang, Yuming and Wang, Jingxiu and Shen, Chenglong and Ye, Pinzhong and Liu, Rui and Chen, Jun and Zhang, Quanhao and Wang, S.},
	month = aug,
	year = {2016},
	pages = {119},
}

@article{Schrijver_2009,
	title = {Driving major solar flares and eruptions: {A} review},
	volume = {43},
	issn = {02731177},
	url = {https://linkinghub.elsevier.com/retrieve/pii/S0273117708005942},
	doi = {10.1016/j.asr.2008.11.004},
	number = {5},
	journal = {AdSpR},
	publisher = {Elsevier},
	author = {Schrijver, Carolus J.},
	month = mar,
	year = {2009},
	pages = {739--755},
}

@article{Lemen_etal_2012,
	title = {The {Atmospheric} {Imaging} {Assembly} ({AIA}) on the {Solar} {Dynamics} {Observatory} ({SDO})},
	volume = {1},
	number = {275},
	journal = {SoPh},
	author = {Lemen, James R and Akin, David J and Boerner, Paul F and Chou, Catherine and Drake, Jerry F and Duncan, Dexter W and Edwards, Christopher G and Friedlaender, Frank M and Heyman, Gary F and Hurlburt, Neal E and {others}},
	year = {2012},
	pages = {17--40},
}

@article{Alissandrakis_1981,
	title = {On the computation of constant {\textbackslash}textbackslashalpha force-free magnetic field},
	volume = {100},
	url = {http://adsabs.harvard.edu/full/1981A%26A...100..197A},
	number = {1},
	journal = {A\&A},
	author = {Alissandrakis, C. E.},
	year = {1981},
	pages = {197--200},
}

@article{Chintzoglou_2018,
	title = {The {Origin} of {Major} {Solar} {Activity}: {Collisional} {Shearing} between {Nonconjugated} {Polarities} of {Multiple} {Bipoles} {Emerging} within {Active} {Regions}},
	volume = {871},
	issn = {1538-4357},
	url = {http://arxiv.org/abs/1811.02186},
	doi = {10.3847/1538-4357/aaef30},
	number = {1},
	journal = {ApJ},
	publisher = {IOP Publishing},
	author = {Chintzoglou, Georgios and Zhang, Jie and Cheung, Mark C.M. M. and Kazachenko, Maria},
	month = jan,
	year = {2019},
	pages = {67},
}

@article{Toriumi_2017,
	title = {Numerical {Simulations} of {Flare}-productive {Active} {Regions}: δ -sunspots, {Sheared} {Polarity} {Inversion} {Lines}, {Energy} {Storage}, and {Predictions}},
	volume = {850},
	issn = {1538-4357},
	url = {http://dx.doi.org/10.3847/1538-4357/aa95c2},
	doi = {10.3847/1538-4357/aa95c2},
	number = {1},
	journal = {ApJ},
	publisher = {IOP Publishing},
	author = {Toriumi, Shin and Takasao, Shinsuke},
	month = nov,
	year = {2017},
	pages = {39},
}

@article{Bobra_2014,
	title = {The {Helioseismic} and {Magnetic} {Imager} ({HMI}) {Vector} {Magnetic} {Field} {Pipeline}: {SHARPs} – {Space}-{Weather} {HMI} {Active} {Region} {Patches}},
	volume = {289},
	issn = {0038-0938},
	url = {http://link.springer.com/10.1007/s11207-014-0529-3},
	doi = {10.1007/s11207-014-0529-3},
	number = {9},
	urldate = {2014-11-18},
	journal = {SoPh},
	publisher = {Springer},
	author = {Bobra, Monica G. and Sun, Xudong and Hoeksema, J. Todd and Turmon, M. and Liu, Yang and Hayashi, Keiji and Barnes, Graham and Leka, K. D.},
	month = sep,
	year = {2014},
	pages = {3549--3578},
}

@article{Wangd_2017,
	title = {Critical {Height} of the {Torus} {Instability} in {Two}-ribbon {Solar} {Flares}},
	volume = {843},
	issn = {2041-8213},
	url = {http://dx.doi.org/10.3847/2041-8213/aa79f0},
	doi = {10.3847/2041-8213/aa79f0},
	number = {1},
	journal = {ApJL},
	publisher = {IOP Publishing},
	author = {Wang, Dong and Liu, Rui and Wang, Yuming and Liu, Kai and Chen, Jun and Liu, Jiajia and Zhou, Zhenjun and Zhang, Min},
	month = jun,
	year = {2017},
	pages = {L9},
}

@article{Kunzel_1960,
	title = {Die {Flare}-{Häufigkeit} in {Fleckengruppen} unterschiedlicher {Klasse} und magnetischer {Struktur}},
	volume = {285},
	url = {http://adsabs.harvard.edu/abs/1960AN....285..271K},
	number = {87},
	journal = {AN},
	author = {Künzel, H},
	year = {1960},
	pages = {271},
}

@article{Liu_2017,
	title = {The causes of quasi-homologous {CMEs}},
	volume = {844},
	copyright = {All rights reserved},
	doi = {10.3847/1538-4357/aa7d56},
	number = {2},
	journal = {ApJ},
	author = {Liu, Lijuan and Wang, Yuming and Liu, Rui and Zhou, Zhenjun and Temmer, M and Liu, Jiajia and Liu, Kai and Shen, Chenglong and Zhang, Quanhao},
	month = aug,
	year = {2017},
	pages = {1--41},
}

@article{Hoeksema_etal_2014,
	title = {The {Helioseismic} and {Magnetic} {Imager} ({HMI}) {Vector} {Magnetic} {Field} {Pipeline}: {Overview} and {Performance}},
	volume = {289},
	doi = {10.1007/s11207-014-0516-8},
	journal = {SoPh},
	author = {Hoeksema, J T and Liu, Yang and Hayashi, Keiji and Sun, Xudong},
	year = {2014},
	pages = {3483--3530},
}

@article{Toriumi_2016,
	title = {{MAGNETIC} {PROPERTIES} {OF} {SOLAR} {ACTIVE} {REGIONS} {THAT} {GOVERN} {LARGE} {SOLAR} {FLARES} {AND} {ERUPTIONS}},
	volume = {834},
	issn = {1538-4357},
	url = {https://iopscience.iop.org/article/10.3847/1538-4357/834/1/56},
	doi = {10.3847/1538-4357/834/1/56},
	number = {1},
	journal = {ApJ},
	publisher = {IOP Publishing},
	author = {Toriumi, Shin and Schrijver, Carolus J. and Harra, Louise K. and Hudson, Hugh and Nagashima, Kaori},
	month = dec,
	year = {2016},
	pages = {56},
}

@article{Leka_Barnes_2003a,
	title = {Photospheric {Magnetic} {Field} {Properties} of {Flaring} versus {Flare}‐quiet {Active} {Regions}. {I}. {Data}, {General} {Approach}, and {Sample} {Results}},
	volume = {595},
	issn = {0004-637X},
	url = {https://iopscience.iop.org/article/10.1086/377511},
	doi = {10.1086/377511},
	number = {2},
	journal = {ApJ},
	publisher = {IOP Publishing},
	author = {Leka, K. D. and Barnes, G.},
	month = oct,
	year = {2003},
	pages = {1277--1295},
}

@article{Falconer_2002,
	title = {Correlation of the {Coronal} {Mass} {Ejection} {Productivity} of {Solar} {Active} {Regions} with {Measures} of {Their} {Global} {Nonpotentiality} from {Vector} {Magnetograms}: {Baseline} {Results}},
	volume = {569},
	issn = {0004-637X},
	url = {https://iopscience.iop.org/article/10.1086/339161},
	doi = {10.1086/339161},
	number = {2},
	journal = {ApJ},
	author = {Falconer, D. A. and Moore, R. L. and Gary, G. A.},
	month = apr,
	year = {2002},
	pages = {1016--1025},
}

@article{Wangym_2011,
	title = {Statistical study of coronal mass ejection source locations: {Understanding} {CMEs} viewed in coronagraphs},
	volume = {116},
	doi = {10.1029/2010JA016101},
	number = {A04104},
	journal = {JGRA},
	publisher = {Wiley Online Library},
	author = {Wang, Yuming and Chen, Caixia and Gui, Bin and Shen, Chenglong and Ye, Pinzhong and Wang, S},
	year = {2011},
}

@article{Chen_Wang_2012,
	title = {Quantifying solar superactive regions with vector magnetic field observations},
	volume = {543},
	issn = {0004-6361},
	url = {http://arxiv.org/abs/1205.6533},
	doi = {10.1051/0004-6361/201118037},
	urldate = {2014-11-18},
	journal = {A\&A},
	author = {Chen, AQ Q and Wang, JX X},
	month = jul,
	year = {2012},
	pages = {A49},
}

@article{Wang_Zhang_2008,
	title = {A {Statistical} {Study} of {Solar} {Active} {Regions} {That} {Produce} {Extremely} {Fast} {Coronal} {Mass} {Ejections}},
	volume = {680},
	issn = {0004-637X},
	url = {http://stacks.iop.org/0004-637X/680/i=2/a=1516},
	doi = {10.1086/587619},
	number = {2},
	journal = {ApJ},
	publisher = {IOP Publishing},
	author = {Wang, Yuming and Zhang, Jie},
	month = jun,
	year = {2008},
	pages = {1516--1522},
}

@article{Liuy_2008,
	title = {Magnetic {Field} {Overlying} {Solar} {Eruption} {Regions} and {Kink} and {Torus} {Instabilities}},
	volume = {679},
	issn = {0004-637X},
	url = {http://stacks.iop.org/1538-4357/679/i=2/a=L151},
	doi = {10.1086/589282},
	number = {2},
	journal = {ApJ},
	publisher = {IOP Publishing},
	author = {Liu, Y},
	month = jun,
	year = {2008},
	pages = {L151--L154},
}

@article{Tian_2002,
	title = {The {Most} {Violent} {Super}-{Active} {Regions} in the 22nd and 23rd {Cycles}},
	volume = {209},
	issn = {00380938},
	url = {http://adsabs.harvard.edu/abs/2002SoPh..209..361T},
	doi = {10.1023/A:1021270202680},
	number = {2},
	urldate = {2014-11-18},
	journal = {SoPh},
	publisher = {Springer},
	author = {Tian, Lirong and Liu, Yang and Wang, Jinxiu},
	year = {2002},
	pages = {361--374},
}

@article{Wangym_2007,
	title = {A {Comparative} {Study} between {Eruptive} {X}‐{Class} {Flares} {Associated} with {Coronal} {Mass} {Ejections} and {Confined} {X}‐{Class} {Flares}},
	volume = {665},
	issn = {0004-637X},
	url = {http://stacks.iop.org/0004-637X/665/i=2/a=1428},
	doi = {10.1086/519765},
	number = {2},
	journal = {ApJ},
	publisher = {IOP Publishing},
	author = {Wang, Yuming and Zhang, Jie},
	month = aug,
	year = {2007},
	pages = {1428--1438},
}

@article{Schrijver_2007,
	title = {A characteristic magnetic field pattern associated with all major solar flares and its use in flare forecasting},
	volume = {655},
	number = {2},
	journal = {ApJ},
	publisher = {IOP Publishing},
	author = {Schrijver, Carolus J},
	year = {2007},
	pages = {L117},
}

@article{Welsch_2009,
	title = {What {Is} the {Relationship} {Between} {Photospheric} {Flow} {Fields} and {Solar} {Flares}?},
	volume = {705},
	issn = {0004-637X},
	url = {http://arxiv.org/abs/0905.0529},
	doi = {10.1088/0004-637X/705/1/821},
	number = {1},
	journal = {ApJ},
	author = {Welsch, Brian T. and Li, Yan and Schuck, Peter W. and Fisher, George H.},
	month = nov,
	year = {2009},
	pages = {821--843},
}

@article{Shibata_2011,
	title = {Solar {Flares}: {Magnetohydrodynamic} {Processes}},
	volume = {8},
	issn = {1614-4961},
	url = {http://link.springer.com/10.12942/lrsp-2011-6},
	doi = {10.12942/lrsp-2011-6},
	journal = {LRSP},
	author = {Shibata, Kazunari and Magara, Tetsuya},
	year = {2011},
}

@article{Bobra_Couvidat_2015,
	title = {Solar {Flare} {Prediction} {Using} {SDO} /{HMI} {Vector} {Magnetic} {Field} {Data} {With} a {Machine}-{Learning} {Algorithm}},
	volume = {798},
	issn = {1538-4357},
	url = {http://stacks.iop.org/0004-637X/798/i=2/a=135?key=crossref.909083302f610075f96c5b8c2bc36691},
	doi = {10.1088/0004-637X/798/2/135},
	number = {2},
	journal = {ApJ},
	author = {Bobra, M. G. and Couvidat, S.},
	year = {2015},
	pages = {135},
}

@article{Jing_2006,
	title = {The {Statistical} {Relationship} between the {Photospheric} {Magnetic} {Parameters} and the {Flare} {Productivity} of {Active} {Regions}},
	volume = {644},
	issn = {0004-637X},
	url = {http://stacks.iop.org/0004-637X/652/i=2/a=1796},
	doi = {10.1086/508989},
	number = {2},
	urldate = {2014-11-18},
	journal = {ApJ},
	publisher = {IOP Publishing},
	author = {Jing, Ju and Song, Hui and Abramenko, Valentyna and Tan, Changyi and Wang, Haimin},
	month = dec,
	year = {2006},
	pages = {1796--1796},
}

@article{Kliem_2006,
	title = {Torus {Instability}},
	volume = {96},
	issn = {0031-9007},
	url = {http://link.aps.org/doi/10.1103/PhysRevLett.96.255002},
	doi = {10.1103/PhysRevLett.96.255002},
	number = {25},
	journal = {PhRvL},
	author = {Kliem, B and Török, T},
	year = {2006},
	pages = {255002},
}

@article{Gary_1989,
	title = {{LINEAR} {FORCE}-{FREE} {MAGNETIC} {FIELDS} {FOR} {SOLAR} {EXTRAPOLATION} {AND} {INTERPRETATION}},
	volume = {69},
	urldate = {2022-10-09},
	journal = {ApJS},
	author = {Gary, G Allen},
	year = {1989},
	pages = {323--348},
}

@article{Wheatland_2000,
	title = {The {Origin} of the {Solar} {Flare} {Waiting}-{Time} {Distribution}},
	volume = {536},
	issn = {0004637X},
	url = {https://iopscience.iop.org/article/10.1086/312739},
	doi = {10.1086/312739},
	number = {2},
	journal = {ApJ},
	author = {Wheatland, M. S.},
	month = jun,
	year = {2000},
	pages = {L109--L112},
}

@article{wangym_2002,
	title = {A statistical study on the geoeffectiveness of {Earth}-directed coronal mass ejections from {March} 1997 to {December} 2000},
	volume = {107},
	issn = {0148-0227},
	url = {http://doi.wiley.com/10.1029/2002JA009244},
	doi = {10.1029/2002JA009244},
	number = {A11},
	urldate = {2014-11-18},
	journal = {JGRA},
	publisher = {Wiley Online Library},
	author = {Wang, Y. M. and Ye, P Z and Wang, S and Zhou, G P and Wang, J X},
	year = {2002},
	pages = {SSH--2},
}

@article{Yang_2024,
	title = {Why {Could} a {Newborn} {Active} {Region} {Produce} {Coronal} {Mass} {Ejections}?},
	volume = {973},
	copyright = {All rights reserved},
	issn = {0004-637X},
	url = {https://dx.doi.org/10.3847/1538-4357/ad6900},
	doi = {10.3847/1538-4357/ad6900},
	language = {en},
	number = {2},
	urldate = {2024-11-09},
	journal = {ApJ},
	publisher = {The American Astronomical Society},
	author = {Yang, Hanzhao and Liu, Lijuan},
	month = sep,
	year = {2024},
	pages = {164},
}

@article{Jarolim_2024,
	title = {Magnetic {Field} {Evolution} of the {Solar} {Active} {Region} 13664},
	volume = {976},
	issn = {2041-8205},
	url = {https://dx.doi.org/10.3847/2041-8213/ad8914},
	doi = {10.3847/2041-8213/ad8914},
	language = {en},
	number = {1},
	urldate = {2025-01-03},
	journal = {ApJL},
	publisher = {The American Astronomical Society},
	author = {Jarolim, Robert and Veronig, Astrid M. and Purkhart, Stefan and Zhang, Peijin and Rempel, Matthias},
	month = nov,
	year = {2024},
	pages = {L12},
}

@article{Wang_2024,
	title = {Unveiling key factors in solar eruptions leading to the solar superstorm in 2024 {May}},
	volume = {692},
	copyright = {© The Authors 2024},
	issn = {0004-6361, 1432-0746},
	url = {https://www.aanda.org/articles/aa/abs/2024/12/aa52008-24/aa52008-24.html},
	doi = {10.1051/0004-6361/202452008},
	language = {en},
	urldate = {2025-01-03},
	journal = {A\&A},
	publisher = {EDP Sciences},
	author = {Wang, Rui and Liu, Ying D. and Zhao, Xiaowei and Hu, Huidong},
	month = dec,
	year = {2024},
	pages = {A112},
}

@article{Kontogiannis_2024a,
	title = {The extremely strong non-neutralized electric currents of the unique solar active region {NOAA} 13664},
	volume = {690},
	copyright = {https://creativecommons.org/licenses/by/4.0},
	issn = {0004-6361, 1432-0746},
	url = {https://www.aanda.org/10.1051/0004-6361/202451627},
	doi = {10.1051/0004-6361/202451627},
	language = {en},
	urldate = {2025-01-03},
	journal = {A\&A},
	author = {Kontogiannis, I.},
	month = oct,
	year = {2024},
	pages = {L10},
}

@article{Romano_2024,
	title = {Analyzing the {Sequence} of {Phases} {Leading} to the {Formation} of the {Active} {Region} 13664, with {Potential} {Carrington}-like {Characteristics}},
	volume = {973},
	issn = {2041-8205, 2041-8213},
	url = {https://iopscience.iop.org/article/10.3847/2041-8213/ad77cb},
	doi = {10.3847/2041-8213/ad77cb},
	language = {en},
	number = {1},
	urldate = {2025-01-03},
	journal = {ApJL},
	author = {Romano, P. and Elmhamdi, A. and Marassi, A. and Contarino, {And} L.},
	month = sep,
	year = {2024},
	pages = {L31},
}

@article{Li_2024,
	title = {Various {Features} of the {X}-class {White}-light {Flares} in {Super} {Active} {Region} {NOAA} 13664},
	volume = {972},
	issn = {2041-8205, 2041-8213},
	url = {https://iopscience.iop.org/article/10.3847/2041-8213/ad6d6c},
	doi = {10.3847/2041-8213/ad6d6c},
	language = {en},
	number = {1},
	urldate = {2025-01-03},
	journal = {ApJL},
	author = {Li, Ying and Liu, Xiaofeng and Jing, Zhichen and Chen, Wei and Li, Qiao and Su, Yang and Song, De-Chao and Ding, M. D. and Feng, Li and Li, Hui and Gan, Weiqun},
	month = sep,
	year = {2024},
	pages = {L1},
}

@article{Hayakawa_2024,
	title = {The {Solar} and {Geomagnetic} {Storms} in {May} 2024: {A} {Flash} {Data} {Report}},
	volume = {979},
	shorttitle = {The {Solar} and {Geomagnetic} {Storms} in {May} 2024},
	url = {https://doi.org/10.3847/1538-4357/ad9335},
	doi = {10.3847/1538-4357/ad9335},
	urldate = {2025-01-06},
	journal = {ApJ},
	author = {Hayakawa, Hisashi and Ebihara, Yusuke and Mishev, Alexander and Koldobskiy, Sergey and Kusano, Kanya and Bechet, Sabrina and Yashiro, Seiji and Iwai, Kazumasa and Shinbori, Atsuki and Mursula, Kalevi and Miyake, Fusa and Shiota, Daikou and Silveira, Marcos V. D. and Stuart, Robert and Oliveira, Denny M. and Akiyama, Sachiko and Ohnishi, Kouji and Ledvina, Vincent and Miyoshi, Yoshizumi},
	month = jan,
	year = {2025},
	pages = {49},
}

@misc{Sun_2024,
	title = {Extraordinary {Magnetic} {Flux} {Emergence} {Rate} {Preceding} the {May} 2024 {Extreme} {Geomagnetic} {Disturbances} {\textbar} {HMI} {Science} {Nuggets}},
	url = {http://hmi.stanford.edu/hminuggets/?p=4216},
	language = {en-US},
	urldate = {2025-01-19},
	author = {Sun, Xudong and Norton, Aimee and Shin, Toriumi and Schuck, Peter and Zhang, Jie},
	month = jun,
	year = {2024},
}

@article{Bai_1988,
	title = {Distribution of {Flares} on the {Sun} during 1955--1985: ``{Hot} {Spots}'' ({Active} {Zones}) {Lasting} for 30 {Years}},
	volume = {328},
	issn = {0004-637X},
	shorttitle = {Distribution of {Flares} on the {Sun} during 1955--1985},
	url = {https://ui.adsabs.harvard.edu/abs/1988ApJ...328..860B},
	doi = {10.1086/166344},
	urldate = {2025-02-04},
	journal = {ApJ},
	publisher = {IOP},
	author = {Bai, Taeil},
	month = may,
	year = {1988},
	pages = {860},
}

@article{Hale_1919,
	title = {The {Magnetic} {Polarity} of {Sun}-{Spots}},
	volume = {49},
	issn = {0004-637X},
	url = {https://ui.adsabs.harvard.edu/abs/1919ApJ....49..153H},
	doi = {10.1086/142452},
	urldate = {2025-02-19},
	journal = {ApJ},
	publisher = {IOP},
	author = {Hale, George E. and Ellerman, Ferdinand and Nicholson, S. B. and Joy, A. H.},
	month = apr,
	year = {1919},
	pages = {153},
}

@article{Shen_2021,
	title = {Origin of {Extremely} {Intense} {Southward} {Component} of {Magnetic} {Field} ({Bs}) in {ICMEs}},
	volume = {9},
	issn = {2296-424X},
	url = {https://www.frontiersin.org/journals/physics/articles/10.3389/fphy.2021.762488/full},
	doi = {10.3389/fphy.2021.762488},
	language = {English},
	urldate = {2025-02-20},
	journal = {FrPhy},
	publisher = {Frontiers},
	author = {Shen, Chenglong and Chi, Yutian and Xu, Mengjiao and Wang, Yuming},
	month = nov,
	year = {2021},
}

@article{Camilla_2020,
	title = {{CME}–{CME} {Interactions} as {Sources} of {CME} {Geoeffectiveness}: {The} {Formation} of the {Complex} {Ejecta} and {Intense} {Geomagnetic} {Storm} in 2017 {Early} {September}},
	volume = {247},
	issn = {0067-0049},
	shorttitle = {{CME}–{CME} {Interactions} as {Sources} of {CME} {Geoeffectiveness}},
	url = {https://dx.doi.org/10.3847/1538-4365/ab6216},
	doi = {10.3847/1538-4365/ab6216},
	language = {en},
	number = {1},
	urldate = {2025-02-20},
	journal = {ApJS},
	publisher = {The American Astronomical Society},
	author = {Scolini, Camilla and Chané, Emmanuel and Temmer, Manuela and Kilpua, Emilia K. J. and Dissauer, Karin and Veronig, Astrid M. and Palmerio, Erika and Pomoell, Jens and Dumbović, Mateja and Guo, Jingnan and Rodriguez, Luciano and Poedts, Stefaan},
	month = feb,
	year = {2020},
	pages = {21},
}

@article{Boffetta_1999,
	title = {Power {Laws} in {Solar} {Flares}: {Self}-{Organized} {Criticality} or {Turbulence}?},
	volume = {83},
	copyright = {http://link.aps.org/licenses/aps-default-license},
	issn = {0031-9007, 1079-7114},
	shorttitle = {Power {Laws} in {Solar} {Flares}},
	url = {https://link.aps.org/doi/10.1103/PhysRevLett.83.4662},
	doi = {10.1103/PhysRevLett.83.4662},
	language = {en},
	number = {22},
	urldate = {2025-02-27},
	journal = {PhRvL},
	author = {Boffetta, Guido and Carbone, Vincenzo and Giuliani, Paolo and Veltri, Pierluigi and Vulpiani, Angelo},
	month = nov,
	year = {1999},
	pages = {4662--4665},
}

@article{Jaswal_2025,
	title = {Deconstructing the {Properties} of {Solar} {Super} {Active} {Region} 13664 in the {Context} of the {Historic} {Geomagnetic} {Storm} of 2024 {May} 10–11},
	volume = {979},
	issn = {0004-637X},
	url = {https://dx.doi.org/10.3847/1538-4357/ad960b},
	doi = {10.3847/1538-4357/ad960b},
	language = {en},
	number = {1},
	urldate = {2025-02-27},
	journal = {ApJ},
	publisher = {The American Astronomical Society},
	author = {Jaswal, Priyansh and Sinha, Suvadip and Nandy, Dibyendu},
	month = jan,
	year = {2025},
	pages = {31},
}

@article{Dhakal_2024,
	title = {What {Are} the {Causes} of {Super} {Activity} of {Solar} {Active} {Regions}?},
	volume = {960},
	issn = {0004-637X},
	url = {https://dx.doi.org/10.3847/1538-4357/ad07d2},
	doi = {10.3847/1538-4357/ad07d2},
	language = {en},
	number = {1},
	urldate = {2025-03-09},
	journal = {ApJ},
	publisher = {The American Astronomical Society},
	author = {Dhakal, Suman K. and Zhang, Jie},
	month = dec,
	year = {2023},
	pages = {36},
}

@article{Shen_2018,
	title = {Why the {Shock}-{ICME} {Complex} {Structure} {Is} {Important}: {Learning} from the {Early} 2017 {September} {CMEs}},
	volume = {861},
	issn = {0004-637X},
	shorttitle = {Why the {Shock}-{ICME} {Complex} {Structure} {Is} {Important}},
	url = {https://dx.doi.org/10.3847/1538-4357/aac204},
	doi = {10.3847/1538-4357/aac204},
	language = {en},
	number = {1},
	urldate = {2025-03-14},
	journal = {ApJ},
	publisher = {The American Astronomical Society},
	author = {Shen, Chenglong and Xu, Mengjiao and Wang, Yuming and Chi, Yutian and Luo, Bingxian},
	month = jun,
	year = {2018},
	pages = {28},
}

@article{Chi_2016,
	title = {Statistical {Study} of the {Interplanetary} {Coronal} {Mass} {Ejections} from 1995 to 2015},
	volume = {291},
	issn = {1573-093X},
	url = {https://doi.org/10.1007/s11207-016-0971-5},
	doi = {10.1007/s11207-016-0971-5},
	language = {en},
	number = {8},
	urldate = {2025-03-14},
	journal = {SoPh},
	author = {Chi, Yutian and Shen, Chenglong and Wang, Yuming and Xu, Mengjiao and Ye, Pinzhong and Wang, Shui},
	month = oct,
	year = {2016},
	pages = {2419--2439},
}

@article{Liuyd_2014,
	title = {Observations of an extreme storm in interplanetary space caused by successive coronal mass ejections},
	volume = {5},
	copyright = {2014 Springer Nature Limited},
	issn = {2041-1723},
	url = {https://www.nature.com/articles/ncomms4481},
	doi = {10.1038/ncomms4481},
	language = {en},
	number = {1},
	urldate = {2025-03-14},
	journal = {NatCo},
	publisher = {Nature Publishing Group},
	author = {Liu, Ying D. and Luhmann, Janet G. and Kajdič, Primož and Kilpua, Emilia K. J. and Lugaz, Noé and Nitta, Nariaki V. and Möstl, Christian and Lavraud, Benoit and Bale, Stuart D. and Farrugia, Charles J. and Galvin, Antoinette B.},
	month = mar,
	year = {2014},
	pages = {3481},
}

@article{Bobra_Ilonidis_2016,
	title = {Predicting {Coronal} {Mass} {Ejections} {Using} {Machine} {Learning} {Methods}},
	volume = {821},
	issn = {0004-637X},
	url = {https://doi.org/10.3847/0004-637X/821/2/127},
	doi = {10.3847/0004-637X/821/2/127},
	language = {en},
	number = {2},
	urldate = {2025-10-16},
	journal = {ApJ},
	publisher = {The American Astronomical Society},
	author = {Bobra, M. G. and Ilonidis, S.},
	month = apr,
	year = {2016},
	pages = {127},
}

@article{Tziotziou_2012,
	title = {{THE} {MAGNETIC} {ENERGY}–{HELICITY} {DIAGRAM} {OF} {SOLAR} {ACTIVE} {REGIONS}},
	volume = {759},
	issn = {2041-8205},
	url = {https://doi.org/10.1088/2041-8205/759/1/L4},
	doi = {10.1088/2041-8205/759/1/L4},
	language = {en},
	number = {1},
	urldate = {2025-10-16},
	journal = {ApJL},
	publisher = {The American Astronomical Society},
	author = {Tziotziou, Kostas and Georgoulis, Manolis K. and Raouafi, Nour-Eddine},
	month = oct,
	year = {2012},
	pages = {L4},
}

@article{Tziotziou_2013,
	title = {{INTERPRETING} {ERUPTIVE} {BEHAVIOR} {IN} {NOAA} {AR} 11158 {VIA} {THE} {REGION}'{S} {MAGNETIC} {ENERGY} {AND} {RELATIVE}-{HELICITY} {BUDGETS}},
	volume = {772},
	issn = {0004-637X},
	url = {https://doi.org/10.1088/0004-637X/772/2/115},
	doi = {10.1088/0004-637X/772/2/115},
	language = {en},
	number = {2},
	urldate = {2025-10-16},
	journal = {ApJ},
	publisher = {The American Astronomical Society},
	author = {Tziotziou, Kostas and Georgoulis, Manolis K. and Liu, Yang},
	month = jul,
	year = {2013},
	pages = {115},
}

@article{McIntosh_1990,
	title = {The classification of sunspot groups},
	volume = {125},
	copyright = {1990 Kluwer Academic Publishers},
	issn = {1573-093X},
	url = {https://link.springer.com/article/10.1007/BF00158405},
	doi = {10.1007/BF00158405},
	language = {En},
	number = {2},
	urldate = {2025-10-17},
	journal = {SoPh},
	publisher = {Springer},
	author = {McIntosh, Patrick S.},
	month = sep,
	year = {1990},
	pages = {251--267},
}

@article{James_2022,
	title = {Evolution of the critical torus instability height and coronal mass ejection likelihood in solar active regions},
	volume = {665},
	copyright = {© A. W. James et al. 2022},
	issn = {0004-6361},
	url = {https://www.aanda.org/articles/aa/full_html/2022/09/aa42910-21/aa42910-21.html},
	language = {en-gb},
	urldate = {2025-10-17},
	journal = {A\&A},
	author = {James, Alexander W. and Williams, David R. and O’Kane, Jennifer},
	month = jun,
	year = {2022},
	pages = {A37},
}

@article{Gupta_2024,
	title = {Stability of the coronal magnetic field around large confined and eruptive solar flares},
	volume = {686},
	copyright = {© The Authors 2024},
	issn = {0004-6361},
	url = {https://www.aanda.org/articles/aa/full_html/2024/06/aa46212-23/aa46212-23.html},
	language = {en-gb},
	urldate = {2025-10-17},
	journal = {A\&A},
	author = {Gupta, M. and Thalmann, J. K. and Veronig, A. M.},
	month = jun,
	year = {2024},
	pages = {A115},
}

@article{Baumgartner_2018,
	title = {On the {Factors} {Determining} the {Eruptive} {Character} of {Solar} {Flares}},
	volume = {853},
	issn = {0004-637X},
	url = {https://ui.adsabs.harvard.edu/abs/2018ApJ...853..105B},
	doi = {10.3847/1538-4357/aaa243},
	urldate = {2025-10-17},
	journal = {ApJ},
	publisher = {IOP},
	author = {Baumgartner, Christian and Thalmann, Julia K. and Veronig, Astrid M.},
	month = feb,
	year = {2018},
	pages = {105},
}

@article{Filippov_2020,
	title = {Failed prominence eruptions near 24 cycle maximum},
	volume = {494},
	url = {https://dx.doi.org/10.1093/mnras/staa896},
	doi = {10.1093/mnras/staa896},
	language = {en},
	number = {2},
	urldate = {2025-10-17},
	journal = {MNRAS},
	author = {Filippov, B.},
	month = apr,
	year = {2020},
	pages = {2166--2177},
}

@article{Dikpati_2025,
	title = {Mother’s {Day} {Superstorms}: {Pre}- and {Post}-storm {Evolutionary} {Patterns} of {ARs} 13664/8},
	volume = {988},
	issn = {0004-637X},
	shorttitle = {Mother’s {Day} {Superstorms}},
	url = {https://doi.org/10.3847/1538-4357/addd09},
	doi = {10.3847/1538-4357/addd09},
	language = {en},
	number = {1},
	urldate = {2026-03-12},
	journal = {ApJ},
	publisher = {The American Astronomical Society},
	author = {Dikpati, Mausumi and Korsós, Marianna B. and Norton, Aimee A. and Raphaldini, Breno and Jain, Kiran and McIntosh, Scott W. and Gilman, Peter A. and Teruya, Andre S. W. and Raouafi, Nour E.},
	month = jul,
	year = {2025},
	pages = {108},
}

@article{Korsos_2025,
	title = {Tracing {Magnetic} {Signatures} through the {Solar} {Atmosphere} in {Seven} {Eruptive} {Active} {Regions}},
	volume = {990},
	issn = {0004-637X},
	url = {https://doi.org/10.3847/1538-4357/adf2a7},
	doi = {10.3847/1538-4357/adf2a7},
	language = {en},
	number = {2},
	urldate = {2026-03-12},
	journal = {ApJ},
	publisher = {The American Astronomical Society},
	author = {Korsós, Marianna B. and Kontogiannis, Ioannis and Shukhobodskaia, Daria and Rodriguez, Luciano and Ferrente, Fabiana and Zuccarello, Francesca},
	month = sep,
	year = {2025},
	pages = {121},
}

@article{Jing_2025,
	title = {The {M}- and {X}-class {White}-light {Flares} in {Super} {Active} {Region} {NOAA} 13664/13697 {Observed} by {ASO}-{S}/{LST}/{WST}},
	volume = {992},
	issn = {0004-637X},
	url = {https://doi.org/10.3847/1538-4357/ae0708},
	doi = {10.3847/1538-4357/ae0708},
	language = {en},
	number = {1},
	urldate = {2026-03-12},
	journal = {ApJ},
	publisher = {The American Astronomical Society},
	author = {Jing, Zhichen and Li, Ying and Li, Jingwei and Li, Qiao},
	month = oct,
	year = {2025},
	pages = {72},
}

@article{Razquin_2025,
	title = {Coronal dimmings from active region 13664 during the {May} 2024 solar energetic events},
	volume = {699},
	copyright = {https://creativecommons.org/licenses/by/4.0},
	issn = {0004-6361, 1432-0746},
	url = {https://www.aanda.org/10.1051/0004-6361/202554772},
	doi = {10.1051/0004-6361/202554772},
	language = {en},
	urldate = {2026-03-12},
	journal = {A\&A},
	author = {Razquin, Amaia and Veronig, Astrid M. and Dissauer, Karin and Podladchikova, Tatiana and Jain, Shantanu},
	month = jul,
	year = {2025},
	pages = {A40},
}

@article{Zuccarello_2014,
	title = {{OBSERVATIONAL} {EVIDENCE} {OF} {TORUS} {INSTABILITY} {AS} {TRIGGER} {MECHANISM} {FOR} {CORONAL} {MASS} {EJECTIONS}: {THE} 2011 {AUGUST} 4 {FILAMENT} {ERUPTION}},
	volume = {785},
	issn = {0004-637X},
	shorttitle = {{OBSERVATIONAL} {EVIDENCE} {OF} {TORUS} {INSTABILITY} {AS} {TRIGGER} {MECHANISM} {FOR} {CORONAL} {MASS} {EJECTIONS}},
	url = {https://doi.org/10.1088/0004-637X/785/2/88},
	doi = {10.1088/0004-637X/785/2/88},
	language = {en},
	number = {2},
	urldate = {2026-03-12},
	journal = {ApJ},
	publisher = {The American Astronomical Society},
	author = {Zuccarello, F. P. and Seaton, D. B. and Mierla, M. and Poedts, S. and Rachmeler, L. A. and Romano, P. and Zuccarello, F.},
	month = mar,
	year = {2014},
	pages = {88},
}

@article{MacTaggart_2025,
	title = {The {Magnetic} {Topology} of {AR13664} {Leading} to {Its} {First} {Halo} {CME}},
	volume = {130},
	copyright = {©2025. The Author(s).},
	issn = {2169-9402},
	url = {https://onlinelibrary.wiley.com/doi/abs/10.1029/2024JA033462},
	doi = {10.1029/2024JA033462},
	language = {en},
	number = {4},
	urldate = {2026-03-22},
	journal = {JGRA},
	author = {MacTaggart, D. and Williams, T. and Aslam, O. P. M.},
	year = {2025},
	pages = {e2024JA033462},
}

@article{McCloskey_2018,
	title = {Flare forecasting using the evolution of {McIntosh} sunspot classifications},
	volume = {8},
	copyright = {© A.E. McCloskey et al., Published by EDP Sciences 2018},
	issn = {2115-7251},
	url = {https://www.swsc-journal.org/articles/swsc/abs/2018/01/swsc170097/swsc170097.html},
	doi = {10.1051/swsc/2018022},
	language = {en},
	urldate = {2026-03-23},
	journal = {JSWSC},
	publisher = {EDP Sciences},
	author = {McCloskey, Aoife E. and Gallagher, Peter T. and Bloomfield, D. Shaun},
	year = {2018},
	pages = {A34},
}
  
\end{document}